%% file: Article-arXiv.tex
\documentclass[10pt, prd, reprint, onecolumn, nofootinbib]{revtex4-2}

\usepackage[utf8]{inputenc}

\usepackage{amsmath}

\usepackage{amsfonts}
\usepackage{mathrsfs}

\usepackage{amssymb}
\usepackage{units}

\usepackage{graphicx}

\usepackage{array}

\usepackage{placeins}

\usepackage{verbatim}

\usepackage{accents}

\usepackage{multirow}

\usepackage{color}

\newcommand{\mrm}{\mathrm}

\makeatletter

\let\old@dmathbeg\[
\let\old@dmathend\]

\newcommand{\rovnec}[1]{\old@dmathbeg#1\old@dmathend}
\newcommand{\rovcis}[2]{\begin{equation}#1\label{#2}\end{equation}}
\newcommand{\drovcis}[2]{\begin{equation}\begin{split}#1\end{split}\label{#2}\end{equation}}
\newcommand{\drovnec}[1]{\begin{equation*}\begin{split}#1\end{split}\end{equation*}} 
\newcommand{\provcis}[1]{\begin{align}#1\end{align}}
\newcommand{\provnec}[1]{\begin{align*}#1\end{align*}}

\newcommand{\rov}{\@ifstar\rovnec\rovcis}
\newcommand{\drov}{\@ifstar\drovnec\drovcis}
\newcommand{\prov}{\@ifstar\provnec\provcis}

\newcommand{\vast}{\bBigg@{4}}
\newcommand{\Vast}{\bBigg@{5}}

\newcommand{\setstretch}[1]{%
  \def\baselinestretch{#1}%
  \@currsize
}

\makeatother

\DeclareMathOperator{\sgn}{sgn}

\DeclareMathOperator{\diffbold}{\mathbf{d}}
\newcommand{\bd}{\diffbold\!}

\newcommand{\mca}{\mathcal}

\newcommand{\mbs}{\boldsymbol}

\DeclareFontEncoding{LGR}{}{}

\DeclareMathAlphabet{\msi}{OT1}{cmss}{m}{it}

\DeclareMathAlphabet{\mgr}{LGR}{cmr}{m}{n}

\newcommand{\rpi}{\mgr{p}}

\renewcommand{\[}{\left[}
\renewcommand{\]}{\right]}

\newcommand{\f}{\!\left}
\newcommand{\ri}{\right}

\newcommand{\pder}[3][]{\frac{\partial^{#1}#2}{\partial{#3}^{#1}}}
\newcommand{\dpder}[2]{\pder[2]{#1}{#2}}

\newcommand{\res}[2]{\left.#1\ri|_{#2}}
\newcommand{\bres}[2]{#1\big|_{#2}}
\newcommand{\lbl}{\label}
\newcommand{\rvt}{\ .}
\newcommand{\rvc}{\ ,}
\newcommand{\rvs}{\ ;}

\renewcommand{\(}{\left(}
\renewcommand{\)}{\right)}

\begin{document}

\title{Extraction of energy from an extremal rotating electrovacuum black hole:\\ Particle collisions in the equatorial plane}

\author{Filip Hejda}
\email{hejdaf@fzu.cz}
\affiliation{CEICO, Institute of Physics of the Czech Academy of Sciences, Na Slovance 1999/2, 182 21 Prague 8, Czech Republic}
\affiliation{Centro de Astrofísica e
Gravitação -- CENTRA, Departamento de
Física, Instituto Superior Técnico -- IST, Universidade de Lisboa -- UL, Avenida Rovisco Pais 1, 1049-001 Lisboa, Portugal}
\author{José P. S. Lemos}
\email{joselemos@ist.utl.pt}
\affiliation{Centro de Astrofísica e
Gravitação  -- CENTRA, Departamento de
Física, Instituto Superior Técnico -- IST, Universidade de Lisboa -- UL, Avenida Rovisco Pais 1, 1049-001 Lisboa, Portugal}
\author{Oleg B. Zaslavskii}
\email{zaslav@ukr.net}
\affiliation{Department of Physics and Technology,
Kharkov V. N. Karazin National University,
4 Svoboda Square, Kharkov 61022, Ukraine
}
\affiliation{
Institute of Mathematics and Mechanics,
Kazan Federal University, 18 Kremlyovskaya Street,
Kazan 420008, Russia}

\begin{abstract}
The collisional Penrose process received much attention when Bañados, Silk
and West (BSW) pointed out the possibility of test-particle collisions
with arbitrarily high center-of-mass energy in the vicinity of the
horizon of an extremally rotating black hole. However, the energy that
can be extracted from the black hole in this promising, if simplified,
scenario, called the BSW effect, turned out to be subject to unconditional
upper bounds. And although such bounds were not found for the
electrostatic variant of the process, this version is also
astrophysically unfeasible, since it requires a maximally charged
black hole.
In order to deal with these deficiencies, we revisit the unified version
of the BSW effect concerning collisions
of charged particles in the equatorial plane of a rotating
electrovacuum black hole spacetime. Performing a general analysis of
energy extraction through this process, we explain in detail how the
seemingly incompatible limiting cases arise. Furthermore, we
demonstrate that the unconditional upper bounds on the extracted
energy are absent for arbitrarily small values of the black hole
electric charge.
Therefore, our setup represents an intriguing simplified model for
possible highly energetic processes happening around astrophysical
black holes, which may spin fast but can have only a tiny electric charge
induced via interaction with an external magnetic field.
\end{abstract}

\maketitle


\section{Introduction}

Penrose \cite{Penrose69} proposed a mechanism to extract energy from a rotating vacuum black hole through a test particle disintegration in its vicinity; one fragment can escape with more energy than the energy
the original particle had, if the other fragment falls inside the black hole and reduces slightly its angular momentum \cite{Christd70}. However, serious doubts about the practical relevance of this original variant of the Penrose process were raised early on \cite{BarPrTeu, Wald74a}. 
A major obstacle is the fact that the fragments need to have relative velocity of more than half the speed of light, which is very restrictive. Nevertheless, such issues can be resolved by considering more general variants of the process. A key ingredient for one of the remedies was provided by Wald \cite{Wald74}, who realized that a rotating black hole in an external magnetic field can become charged due to selective charge accretion.
Using Wald's weak-field solution as background,
particle disintegration into oppositely charged fragments
was considered, and it was shown 
that the requirement of high relative velocity can be circumvented \cite{WaDhuDa}. This generalization, or revival, as the authors put it, of the Penrose process can be also understood as a crossover between its original variant and its electrostatic version described for nonrotating black holes \cite{DenaRuff}.

Another way of fixing the shortcomings of the original Penrose process is to consider particle collisions instead of decays. The required high relative velocity of the final particles can then arise naturally as a result of a high-energy collision. Interestingly, it has been noted that the relative Lorentz factor can in fact diverge in some cases, if the collision point is taken toward the horizon. 
 In particular, this happens
for a collision between an orbiting and an infalling particle in the case of an extremal black hole \cite{PirShK}
and also
for a collision between a radially outgoing and a radially incoming
particle \cite{PirSh}.
Neither of these options seems very realistic, as both involve
particles confined to the vicinity of the horizon. However, whereas
the latter one generically requires a white hole horizon as explained
in \cite{Zasl16}, the former one has viable variants. Notably, Bañados,
Silk, and West discovered its modification with both particles coming
from rest at infinity \cite{BSW}. This BSW effect requires a
fine-tuned particle, called critical particle, which can only
asymptotically approach the horizon radius, as if approaching an orbit
(see, e.g., the discussion in Section IV B in \cite{a2}). For more types
of near-horizon high-energy collisional processes involving orbiting
particles, see, e.g., \cite{HK11a, Zasl12d}. A broad overview of the collisional Penrose process and the BSW effect covering many additional aspects can be found in the work of Schnittman \cite{Schnitt18}.

Since the BSW effect has been derived in the test particle approximation and it relies on fine tuning and extremality of the black hole, there has been an actual concern that it may get suppressed in more realistic circumstances \cite{BCGPS} 
(see also \cite{HK14} for a review).
 But, surprisingly, it turned out that the energy extraction is quite unsatisfactory even with all the simplifying assumptions in place. Namely, it has been established almost simultaneously both by numerical \cite{BPAH} and analytical means \cite{HaNeMi} (see also \cite{Zasl12b}) that there is an unconditional upper bound on the extracted energy despite the center-of-mass collision energy being unbounded. 
 Remarkably enough, Schnittman \cite{Schnitt14} discovered a scenario that is more favorable for energy extraction than the BSW effect with its precise fine tuning. 
    A nearly critical particle, i.e, a particle
    with imperfect fine tuning, can turn from incoming to outgoing motion in the radial direction before colliding with another particle in the vicinity of the black hole, which is advantageous. Nevertheless, as further clarified by additional analytical studies \cite{OHM, Zasl16b}, the enhancement only consists in replacing one unconditional upper bound on the extracted energy with another, higher one. Let us note that for vacuum spacetimes, such limitations can be overcome, if one considers more general objects than black holes, e.g., naked singularities, see \cite{PHNJK, TanatZasl17}. For different ways to examine the original BSW effect with improved realism, see \cite{TanatZasl13, LibPfeRel}.

Similarly to the original Penrose process, an electrostatic variant of the BSW effect exists for maximally charged, and so nonrotating, black holes \cite{Zasl11a}; it requires fine-tuned charged particles. Surprisingly, no unconditional upper bounds on the extracted energy were found in this case \cite{Zasl12c}. 
Given this, it is natural to ask whether similar results can be obtained in a more realistic situation with arbitrarily small black hole charge. 
One such possibility is the simple case of charged particles moving along the axis of symmetry of a rotating electrovacuum black hole, which was considered in \cite{a3}. Although it was confirmed that there is no upper bound on the extracted energy regardless of how small the black hole charge might be, several caveats were found to make this setup unfeasible for microscopic particles. This motivates us to turn to the more complicated case of collisions of charged particles in the equatorial plane of a rotating electrovacuum black hole. 
Such a crossover between the original version and the electrostatic variant of the BSW effect has been considered in \cite{a2}, yet concerning only
what happens before the particle collision, i.e., 
the approach phase of the process. In the present paper, we shall study energy extraction in this setup; 
 let us emphasize that the key innovation in our discussion here consists in taking into account the simultaneous influence of rotation and electric charge. 
Our main purpose is to show that in this case there is no
unconditional upper bound on the extracted energy whenever both the
black hole and the escaping particles are charged.
In our analysis, we draw on some additional works 
\cite{Zasl15b,Zasl10,Zasl15a,HK11b,Article1}.

In the context of this paper, let us mention that black holes with
nonnegligible electric
charge have recently seen renewed interest, as they can play
a role in the mechanism behind fast radio bursts (FRBs).  For example,
it was suggested that a merger of black holes, at least one of which
has enough
 electric charge, can produce FRBs due to the rapidly changing
magnetic dipole moment \cite{Zhang16}. A further study \cite{LiuRomeroLiuLi} investigated the possible role of magnetospheric instability in producing FRBs, both for isolated Kerr-Newman black holes and for binaries.
Yet another, more conventional
model describes how a FRB can result from a prompt discharge of a
metastable collapsed state of a Kerr-Newman black hole
\cite{PunslyBini}. Formation of Kerr-Newman black holes through
collapse of rotating and magnetized neutron stars has been
systematically studied in \cite{NaMoRe17}.

The paper is organized as follows.
In Sec.~\ref{sek:egm}, we describe the properties of motion of charged
test particles around an electrovacuum black hole and classify the
types of motion near an extremal horizon. We also give formulas for
the collision energy in the center-of-mass frame, which diverges in
the horizon limit when one of the particles is critical.
In Sec.~\ref{sek:ap}, we discuss restrictions on the parameters of
critical particles that can be involved in near-horizon high-energy
collisions.  In particular, we determine possible bounds on these
parameters.
In Sec.~\ref{sek:extra}, we perform a full
analysis for energy
extraction. We consider different kinematic regimes, in which
particles can be produced in near-horizon high-energy collisions, and
determine which ones allow the particles to escape. Then, we study
bounds on parameters of the escaping particles; we put emphasis on
situations in which the energy of escaping particles is not bounded.
The results are derived using a general metric form, which makes them
valid also for dirty black holes, i.e., those surrounded by
matter. Additionally, we explain how the previously known limiting
cases can be derived from the general case.
In Sec.~\ref{sek:kn}, we apply the general results to the Kerr-Newman
solution so that we can highlight the whole method using
relevant figures.
In Sec.~\ref{sek:concl}, we conclude.

\section{Motion and collisions of charged test particles}

\lbl{sek:egm}

\subsection{Spacetime metric and electromagnetic potential}

We shall consider a general stationary, axially symmetric spacetime representing an isolated black hole,
with metric $\mbs g$ in coordinates $(t,\varphi,r,\vartheta)$ given by
\rov{\mbs g=-N^2\bd t^2+g_{\varphi\varphi}\(\bd\varphi-\omega\bd t\)^2+g_{rr}\bd r^2+g_{\vartheta\vartheta}\bd\vartheta^2\rvt}{axst}
Here, $N^2$ is the lapse function;
$g_{\varphi\varphi}$,
$g_{rr}$,
$g_{\vartheta\vartheta}$ are the respective metric potentials; 
and $\omega$ is the dragging potential.
We assume $g_{\varphi\varphi}>0$, and also that the product $N\sqrt{g_{rr}}>0$
is finite and nonvanishing 
even for $N\to0$.

Let us further assume that our spacetime is permeated by a Maxwell field obeying the same symmetry as the metric \eqref{axst}. 
We fix the gauge for its potential $\mbs A$ to manifest this symmetry, namely, 
${\mbs A}=A_t\bd t+A_\varphi\bd\varphi$, or rearranging,
\rov{\mbs A=-\phi\bd t+A_\varphi\(\bd\varphi-\omega\bd t\)\rvt}{aaxst}
The component \rov{\phi=-A_t-\omega A_\varphi}{phigen} is called the generalized electrostatic potential. 

\subsection{General equations of equatorial motion}

Let us now consider the motion of test particles with rest mass $m$
and electric charge $q$ in the spacetime defined in Eq.~\eqref{axst}.
Because of the two symmetries that we assumed, there exist two
quantities that are conserved during the electrogeodesic motion.
They are 
the energy $E$ and the axial angular momentum
$L_z$ of the test particle.
We also assume that the metric
\eqref{axst} and the electromagnetic field are symmetric with respect
to the reflections $\vartheta\to\rpi-\vartheta$. Then, we can consider
motion confined to the invariant hypersurface
$\vartheta=\frac{\rpi}{2}$, the equatorial plane. For equatorial
particles, $L_z$ is the total angular momentum; hence, we can drop the
subscript and write $L\equiv L_z$.
The energy  and the axial angular momentum are given by
\prov{E&=-p_t-qA_t\rvc&L&=p_\varphi+qA_\varphi\rvc\lbl{constm}}
where $p_t$ and $p_\varphi$ are the time and azimuthal components of
the particle's 4-momentum $p_\alpha$.

Defining two auxiliary functions $\mca X$ and $\mca Z$ by 
\prov{\mca X&=E-\omega L-q\phi\rvc&\mca Z=\sqrt{\mca X^2-N^2\[m^2+\frac{\(L-qA_\varphi\)^2}{g_{\varphi\varphi}}\]}\rvc
\lbl{aux1}
}
we can write the contravariant components of the particle's 4-momentum
$p^\alpha$
in a compact form:
\prov{p^t&=\frac{\mca X}{N^2}\rvc&p^\varphi&=\frac{\omega\mca X}{N^2}+\frac{L-qA_\varphi}{g_{\varphi\varphi}}\rvc&p^r={\frac{\sigma\mca Z}{N\sqrt{g_{rr}}}}\rvt\lbl{eomp}}
The parameter $\sigma$
has values 
$\sigma=\pm1$ which determine the direction of the radial motion.
In order for the motion to be allowed, the quantity $\mca Z$ has to be real. 
Outside of the black hole, where $N^2>0$, the condition $\mca Z^2>0$ can be equivalently  stated as
$\left|\mca X\ri|\geqslant N\sqrt{m^2+\frac{\(L-qA_\varphi\)^2}{g_{\varphi\varphi}}}$.
It can be seen that there are two disjoint domains
of allowed motion, one with $\mca X>0$ and the other with $\mca X<0$. These two domains touch
 for $N\to0$, where $\mca X\to0$ becomes possible. However, to preserve causality, we need to enforce $p^t>0$, and thus we restrict to the $\mca X>0$ domain. Then, the requirement for the motion to be allowed becomes
\rov{\mca X\geqslant N\sqrt{m^2+\frac{\(L-qA_\varphi\)^2}{g_{\varphi\varphi}}}\rvt}{mallow}
The lower bound, i.e., the equality of Eq.~\eqref{mallow}, 
${\mca X=N\sqrt{m^2+\frac{\(L-qA_\varphi\)^2}{g_{\varphi\varphi}}}}$,
is the condition for a turning point.

The number of relevant parameters can be reduced depending on whether the particle in question is massive or massless. 
Kinematics of massive particles is determined by three parameters: specific energy $\varepsilon\equiv\frac{E}{m}$, specific angular momentum $l\equiv\frac{L}{m}$, and specific charge $\tilde q\equiv\frac{q}{m}$. 
Kinematics of massless particles, which are electrically neutral, e.g., photons, is characterized solely by the impact parameter defined as $b\equiv\frac{L}{E}$.
Based on the distinction
between massive and massless, additional features of the motion, like the existence of circular  orbits, can be deduced. Effective potentials are frequently employed, both for massive particles
and for massless particles
(see \cite{a2} and references therein for the massive particle case).

\subsection{Near-horizon expansions}

\lbl{odd:nhcr}

We wish to study collisions of particles near the black hole horizon, where $N\to0$. 
Let us denote the values of the various quantities on the outer black hole horizon by a subscript
or a superscript $\mrm{H}$ depending on the
convenience. 
As we consider solely equatorial motion, all quantities in the following are understood to be evaluated at $\vartheta=\frac{\rpi}{2}$, which will not be marked explicitly for brevity. For example, by $A_\varphi^\mrm{H}$, we mean the value of $A_\varphi$ on the horizon at $\vartheta=\frac{\rpi}{2}$.

In the vicinity of the horizon, we can perform expansions in variable
$\(r-r_\mrm{H}\)\ll r_\mrm{H}$.  We are interested in extremal black
holes; the horizon located at $r_\mrm{H}$ is understood to be
degenerate hereafter. Let us expand the dragging potential $\omega$
and the generalized electrostatic potential $\phi$ in first order as
follows,
\prov{
\omega&=\omega_\mrm{H}+\hat\omega\(r-r_\mrm{H}\)+\dots
\rvc&
\phi&=\phi_\mrm{H}+\hat\phi\(r-r_\mrm{H}\)+\dots
\rvc}
respectively, and where $\hat\omega=
\bres{\pder{\omega}{r}}{r=r_\mrm{H}}$
and 
$\hat\phi=\bres{\pder{\phi}{r}}{r=r_\mrm{H}}$.
For extremal black holes, we can also renormalize the lapse function as 
$N^2=\(r-r_\mrm{H}\)^2{\mca N}^2$, which leads, in particular, to
\rov{
{\mca N}^2_\mrm{H}
=\frac{1}{2}\res{\dpder{N^2}{r}}{r=r_\mrm{H}}\rvt}{}
Finally, let us introduce a new set of constants
$\mca X_\mrm{H}$, $\chi$, and $\lambda$, which are preserved throughout
the motion and useful to describe the kinematics of particles close to $r_\mrm{H}$.
They are defined in terms of $E,L,q$ as follows:
\prov{\mca X_\mrm{H}&=E-\omega_\mrm{H}L-q\phi_\mrm{H}\rvc&
\chi&=-\hat\omega 
L-q\hat\phi\rvc
&\lambda\equiv p_\varphi^\mrm{H}&=L-qA_\varphi^\mrm{H}\rvt\lbl{nhpar}}
Note that the
parameter $\lambda$ is now defined in a slightly different way than in the previous paper \cite{a2} on
   charged particle collisions.
The formulas given here can be recast into the convention used in \cite{a2} by putting $\lambda\to-m\lambda A_\varphi^\mrm{H}$.
Two of the new parameters, namely, $\mca X_\mrm{H}$ and $\chi$,
 are expansion coefficients of the function $\mca X$
given in Eq.~\eqref{aux1}, i.e.,
\rov{\mca X\approx\mca X_\mrm{H}+\chi\(r-r_\mrm{H}\)+\dots}{fwnh}
The parameters $E, L, q$ can be expressed in terms of the new ones through inverse relations
\prov{E&=\mca X_\mrm{H}+\frac{\(\omega_\mrm{H}\hat\phi-\hat\omega\phi_\mrm{H}\)
\lambda+\chi A_t^\mrm{H}}{\hat\phi+\hat\omega A_\varphi^\mrm{H}}\rvc
&L&=\frac{\hat\phi\lambda-\chi A_\varphi^\mrm{H}}{\hat\phi+\hat\omega 
A_\varphi^\mrm{H}}\rvc
&q&=-\frac{\hat\omega\lambda+\chi }{\hat\phi+\hat\omega 
A_\varphi^\mrm{H}}
\rvc\lbl{nhparinv}}
which all contain the same expression in the denominator.
Hence, when it vanishes, i.e.,
\rov{\hat\phi+\hat\omega A_\varphi^\mrm{H}=0\rvc}{lindeg}
there is clearly a problem with the definitions given in Eq.~\eqref{nhpar}.
Indeed, if Eq.~\eqref{lindeg} holds, $\chi$ and $\lambda$ become proportional to each other,
$\chi=-\hat\omega\lambda$, and thus the variables $\mca X_\mrm{H}, \chi, \lambda$ no longer span the whole parameter space. 
When this degeneracy happens, we can use $\mca X_\mrm{H},\lambda,q$ as our alternative set of parameters. Then, the inverse relations to express $E, L$ become 
\prov{E&=\mca X_\mrm{H}+\omega_\mrm{H}\lambda-qA_t^\mrm{H}&L&=\lambda +qA_\varphi^\mrm{H}\rvt\lbl{nhparinvdeg}}

The behavior of particles close to the horizon radius $r_\mrm{H}$ depends significantly on the value of $\mca X_\mrm{H}$. 
For particles with $\mca X_\mrm{H}<0$, the condition \eqref{mallow} is necessarily violated near the horizon, and thus the particles cannot get arbitrarily close to $r_\mrm{H}$.
On the other hand, particles with $\mca X_\mrm{H}\geqslant0$ can exist arbitrarily close to $r_\mrm{H}$. Let us discuss these types now.

\subsection{Types of particles close to $r_\mrm{H}$}

\subsubsection{Usual (subcritical) particles}

Particles with $\mca X_\mrm{H}>0$ are bound to fall into the black hole if they move inward and get near the horizon. In our discussion, we will refer to those particles as usual particles.
Let us emphasize that we will not consider outgoing usual particles in the vicinity of $r_\mrm{H}$, since it can be shown that such particles cannot be produced in (generic) near-horizon collisions;
see \cite{Zasl15b}.  In our analysis, we exclude the white hole region from which outgoing usual particles could naturally emerge.
For usual particles approaching $r_\mrm{H}$, the function $\mca Z$ of Eq.~\eqref{aux1}
can be expanded in terms of $N^2$, and consequently of $\mca X$, as follows:
\rov{\mca Z\approx\mca X-\frac{N^2}{2\mca X}\[m^2+\frac{\(L-qA_\varphi\)^2}{g_{\varphi\varphi}}\]+\dots}{znhus}

\subsubsection{Critical particles, especially class I critical particles}

We can also consider particles with $\mca X_\mrm{H}=0$, which are called critical.
They are fine-tuned to be on the verge between not 
being able to reach the horizon and falling into the black hole. 
Here, we use the local notion of critical particles.
For asymptotically flat spacetimes, it is also possible to define the critical particles globally, such that they are on the edge of being able to approach the horizon from infinity; see \cite{BSW}.
 By the definition given in Eq.~\eqref{nhpar}, condition $\mca X_\mrm{H}=0$ 
can be understood also as a constraint for parameters $E,L,q$:
\rov{E-\omega_\mrm{H}L-q\phi_\mrm{H}=0\rvt}{critgen}
The expansion around $r_\mrm{H}$ of
the function $\mca Z$ introduced in Eq.~\eqref{aux1} looks rather different for critical particles,
\rov{\mca Z\approx\sqrt{\chi^2-
{\mca N}^2_\mrm{H}
\(m^2+\frac{\lambda^2}{g_{\varphi\varphi}^\mrm{H}}\)}\(r-r_\mrm{H}\)+\dots}{znhcr}
Let us emphasize that with $\mca X_\mrm{H}=0$ the causality condition $p^t>0$ necessarily implies $\chi>0$.

It can be shown that critical particles cannot approach the horizon unless the black hole is extremal (see, e.g., \cite{Zasl10, a2} and references therein). Harada and Kimura \cite{HK11b} distinguished several subtypes of critical particles, out of which we consider chiefly the class I critical particles. The approximate trajectory of an incoming class I critical particle near $r_\mrm{H}$ has the form $r=r_\mrm{H}\[1+\exp\f(-\frac{\tau}{\tau_\mrm{relax}}\)\]$, where $\tau$ is the proper time and $\tau_\mrm{relax}$ is a positive constant; see \cite{a2} for details. Since critical particles of any type can never reach $r_\mrm{H}$, any collisional process involving them will thus happen at some radius $r_\mrm{C}>r_\mrm{H}$. Therefore, it makes sense to consider particles that behave approximately as critical at a given radius.

\subsubsection{Nearly critical particles}

A particle will behave approximately as critical at a radius $r_\mrm{C}$, if the zeroth-order term
in the expansion of $\mca X$ is of comparable magnitude as the first-order term. To quantify this, let us define a formal expansion,
\rov{\mca X_\mrm{H}\approx-C\(r_\mrm{C}-r_\mrm{H}\)-D\(r_\mrm{C}-r_\mrm{H}\)^2+\dots\rvc}{fwhornc}
where the minus sign
in front of the terms follows the usual convention.
$C$, $D$, and so on, are constants
that
are needed for consistency of momentum conservation law at each expansion order.
However, here, we are interested only in the first order, and so
the constant $C$ is enough for our purposes.

For nearly critical particles, the expansion \eqref{fwnh} evaluated at $r_\mrm{C}$
can be recast using Eq.~\eqref{fwhornc} as 
\rov{\mca X\approx\(\chi-C\)\(r_\mrm{C}-r_\mrm{H}\)+\dots}{fwnhnc}
Similarly, the expansion of function $\mca Z$ defined in Eq.~\eqref{aux1} reads for them
\rov{\mca Z\approx\sqrt{\(\chi-C\)^2-
{\mca N}^2_\mrm{H}
\(m^2+\frac{\lambda^2}{g_{\varphi\varphi}^\mrm{H}}\)}\(r_\mrm{C}-r_\mrm{H}\)+\dots}{znhnc}

Nearly critical particles with $C>0$ cannot fall into the black hole, and they must have a turning point at some
radius smaller than $r_\mrm{C}$. Therefore, it makes sense to study collisional processes near the horizon involving also outgoing nearly critical particles. Furthermore, for particles with  
$\chi\gg C>0$,
we can neglect $C$ and treat them as precisely critical around $r_\mrm{C}$. Thus, we can consider outgoing critical particles, too.

\subsubsection{Class II critical particles and class II nearly critical particles}

There exist values of parameters of
critical particles or nearly
critical particles, for which the leading-order coefficient in the expansion \eqref{znhcr} (or \eqref{znhnc}) vanishes. 
The new leading order then becomes 
\rov{\mca Z\sim\(r_\mrm{C}-r_\mrm{H}\)^\frac{3}{2}}{}
or higher. 
These are the so-called class II critical particles or nearly
critical particles.

Kinematics of class II critical particles represents an interesting theoretical issue, which, however, involves technical complications; cf. \cite{a2}. Moreover, since class II critical particles require fine tuning of
two parameters, instead of just one, they are much less important for practical considerations.
Thus, we will mostly omit details regarding class II critical particles in the following. 

\subsection{BSW effect and its Schnittman variant}

\lbl{odd:BSW}

We have seen that in the near-horizon region of an extremal black hole two distinct types of motion do coexist. Whereas usual particles with $\mca X_\mrm{H}>0$ cross $r_\mrm{H}$ and fall into the black hole, 
critical particles with $\mca X_\mrm{H}=0$ can only approach $r_\mrm{H}$ asymptotically. This leads to a divergent relative Lorentz factor due to the relative velocity between the two types of motion approaching the speed of light.

Hence, the expression for the collision energy in the center-of-mass frame, 
\rov{E^2_\mrm{CM}=m_1^2+m_2^2-2g_{\mu\nu}p_{1}^\mu p_{2}^{\vphantom{\mu}\nu}\rvc}{ecmgen}
will be dominated by the scalar-product term, if we consider near-horizon collisions between critical and usual particles. In particular, inserting \eqref{eomp} for a critical particle labeled $1$ and for a usual particle labeled $2$ and using Eqs.~\eqref{fwnh},
\eqref{znhus}, and \eqref{znhcr},
we find that the leading-order contribution of Eq.~\eqref{ecmgen} is
\rov{E_\mrm{CM}^2\approx\frac{\mca X_2^\mrm{H}}{r_\mrm{C}-r_\mrm{H}}\left
\{\frac{2}{
{\mca N}^2_\mrm{H}}\[\chi_1+\sigma_1\sqrt{\chi_1^2-
{\mca N}^2_\mrm{H}
\(m_1^2+\frac{\lambda_1^2}{g_{\varphi\varphi}^\mrm{H}}\)}\]\ri\}\rvc}{}
see \cite{a2} for details.
A process with an
incoming critical particle ($\sigma_1=-1$) is called BSW type after Bañados, Silk, and West \cite{BSW}, whereas the one with reflected (nearly) critical particle ($\sigma_1=+1$) is called Schnittman type \cite{Schnitt14}. Note that the usual particle is always incoming, i.e., $\sigma_2=-1$. We used the aforementioned approximation $\chi_1\gg C_1>0$ for the Schnittman process.

\section{Kinematics of particles before collision}

\lbl{sek:ap}

\subsection{Admissible region in the parameter space}
\lbl{sek:admreg}

\subsubsection{General considerations}

\lbl{sek:admreg2}

Critical particles are the key ingredient of certain high-energy collisional processes in extremal black hole spacetimes. 
Nevertheless, the parameters of critical particles that can act in such processes are restricted, since the requirement of Eq.~\eqref{mallow} must be fulfilled all the way from the point of inception to the point of collision. 

Let us disregard the concern about where the critical particle originated and focus instead on the point of collision at radius $r_\mrm{C}$. Since we want $r_\mrm{C}$ very close to $r_\mrm{H}$, the 
minimum requirement is that there must be some neighborhood of $r_\mrm{H}$, where condition \eqref{mallow} is satisfied. Using a linear approximation in $r-r_\mrm{H}$, we get the following inequality,
\rov{\chi>\mca N_\mrm{H}\sqrt{m^2+\frac{\lambda^2}{g_{\varphi\varphi}^\mrm{H}}}\rvc}{admreg}
which defines the admissible region of parameters.
Conversely, for parameters satisfying the inequality opposite to Eq.~\eqref{admreg}, condition \eqref{mallow} will be violated in some neighborhood of $r_\mrm{H}$.

Now, the equality 
\rov{\chi=\mca N_\mrm{H}\sqrt{m^2+\frac{\lambda^2}{g_{\varphi\varphi}^\mrm{H}}}}{admhypxlam}
corresponds to the breakdown of the linear approximation of Eq.~\eqref{mallow}. 
Comparing with Eq.~\eqref{znhcr}, we see that Eq.~\eqref{admhypxlam} also implies the critical particles to be class II. Higher-order expansion terms are needed to decide whether motion of class II critical particles is allowed close to $r_\mrm{H}$.
(We note that such higher-order kinematic restrictions were worked out in Sections IV~E and V~B in \cite{a2} and that
additional information on this subject
can be found in  Sec.~II~D and footnote 2 in \cite{a3} and
in Sec.~VII in \cite{Zasl15a}.)
 
Now, let us consider the physical interpretation of the admissible region of parameters. In particular, we would like to distinguish different variants of the collisional processes corresponding to the previously known limiting cases. 
  For extremal vacuum black holes, only critical particles corotating with the black hole can participate in the high-energy collisions, whereas for the nonrotating extremal black holes, the critical particles need to have the same sign of charge as the black hole.
In order to identify counterparts of these limiting variants, which we will call centrifugal mechanism
and electrostatic mechanism, we need to assess how to define the direction in which a charged particle orbits. 

The momentum component $p_\varphi$ determines the direction of motion in $\varphi$ with respect to a locally nonrotating observer (cf. \cite{BarPrTeu}). For uncharged particles, $p_\varphi= L$ is constant, and thus the distinction is universal and unambiguous.  
 Nevertheless, for charged particles, $p_\varphi$ depends on $r$ through the $qA_\varphi$ term.
 Therefore, we essentially need to compare values of $p_\varphi$ at some reference radius.
 A first straightforward choice would be to use $\lambda\equiv p_\varphi^\mrm{H}$;
 see Eq.~\eqref{nhpar}. However, it is clear from Eq.~\eqref{admhypxlam} that one can find points with any value of $\lambda$ in the admissible region (whereas values of $\chi$ in the admissible region are bounded from below by $\chi\geqslant\mca N_\mrm{H}m$). 
Apart from the degenerate case \eqref{lindeg}, no kinematic restriction on $\lambda$ is thus possible. 
Hence, basing the definition of the centrifugal mechanism on $\lambda$ would lead to a trivial result.

A second possible choice is $L$.
What is the justification to use $L$?
We shall consider a region of our spacetime, where the influence of the dragging and of the magnetic field is  insignificant, for example, a far zone of an asymptotically flat spacetime. 
More precisely, let us consider a region where $\omega$ and $A_\varphi$ are negligible, and thus $p_\varphi\approx p^\varphi g_{\varphi\varphi}\approx L$. Then, it readily follows that in such a region particles with $L=0$ move along trajectories of (approximately) constant $\varphi$ and, conversely, particles with different signs of $L$ orbit in different directions therein.
 Hence, we can say that $L$ uniquely distinguishes the direction of motion in $\varphi$ of a particle before it came under the influence of the dragging and of the magnetic field near the black hole. 
 
We conclude that we need to view the admissible region through the parameters $L$ and $q$ for physical interpretation. Similarly to \cite{a2}, let us
focus on Eq.~\eqref{admhypxlam} of the border of the admissible region. Substituting relations \eqref{nhpar} for $\chi$ and $\lambda$ into Eq.~\eqref{admhypxlam} does not generally lead to a single-valued functional dependence between $q, L$. We circumvent this issue by plugging the condition \eqref{admhypxlam} into relations \eqref{nhparinv}, which yields parametric expressions for the border:
\prov{q&=-\frac{\hat\omega\lambda+\mca N_\mrm{H}\sqrt{m^2+\frac{\lambda^2}{g_{\varphi\varphi}^\mrm{H}}}}{\hat\phi+\hat\omega A_\varphi^\mrm{H}}\lbl{qparam}\\L&=\frac{\hat\phi\lambda-\mca N_\mrm{H}A_\varphi^\mrm{H}\sqrt{m^2+\frac{\lambda^2}{g_{\varphi\varphi}^\mrm{H}}}}{\hat\phi+\hat\omega A_\varphi^\mrm{H}}\lbl{Lparam}\\E&=\frac{\(\omega_\mrm{H}\hat\phi-\hat\omega\phi_\mrm{H}\)\lambda+\mca N_\mrm{H}A_t^\mrm{H}\sqrt{m^2+\frac{\lambda^2}{g_{\varphi\varphi}^\mrm{H}}}}{\hat\phi+\hat\omega A_\varphi^\mrm{H}}\rvt\lbl{Eparam}}
Since we are dealing with critical particles, the three expressions are not independent.
In \cite{a2}, different possibilities were distinguished by studying restrictions on signs of
$q$ and $L$ in the admissible region; in particular, the centrifugal mechanism was identified as the case when only the sign of $L$ is restricted, and the electrostatic mechanism was identified as the case when only the sign of $q$ is restricted. 
Here, we employ a complementary, deeper approach and
determine the precise bounds on $q$, $L$, and $E$.

\subsubsection{Bounds on parameters}

\lbl{odd:admbp}

Bounds on values of $q$, $L$, and $E$ in the admissible region will appear as
extrema of expressions \eqref{qparam}-\eqref{Eparam} with respect to $\lambda$.

Let us start with the electric charge $q$.
From Eq.~\eqref{qparam}, we find that a stationary point can occur
at the following 
value of $\lambda$:
\rov{\lambda=-m\frac{g_{\varphi\varphi}^\mrm{H}\hat\omega}{\sqrt{{\mca N}^2_\mrm{H}-g_{\varphi\varphi}^\mrm{H}{\hat\omega}^2}}\rvt}{lamqmin}
Due to the square root in the denominator, 
we need to distinguish three possibilities. 

\noindent
Case {\bf 1a}: If Eq.~\eqref{lamqmin} is imaginary,
Eq.~\eqref{qparam} will take all real values, and hence there is no bound on $q$.

\noindent
Case {\bf 1b}: If Eq.~\eqref{lamqmin} is real, it will correspond to an extremum of Eq.~\eqref{qparam}
with value
\rov{q_\mrm{b}=-\frac{m}{\hat\phi+\hat\omega A_\varphi^\mrm{H}}\sqrt{{\mca N}^2_\mrm{H}-g_{\varphi\varphi}^\mrm{H}{\hat\omega}^2}\rvc}{qmin}
which serves as a bound for $q$.
For Eqs.~\eqref{Lparam} and \eqref{Eparam}, the following values of $L$ and $E$ will be implied by Eq.~\eqref{lamqmin}:
\prov{L&=-\frac{m}{\hat\phi+\hat\omega A_\varphi^\mrm{H}}\frac{{\mca N}^2_\mrm{H}A_\varphi^\mrm{H}+g_{\varphi\varphi}^\mrm{H}\hat\omega\hat\phi}{\sqrt{{\mca N}^2_\mrm{H}-g_{\varphi\varphi}^\mrm{H}{\hat\omega}^2}}\rvc\lbl{qbL}\\E&=\frac{m}{\hat\phi+\hat\omega A_\varphi^\mrm{H}}\frac{{\mca N}^2_\mrm{H}A_t^\mrm{H}-g_{\varphi\varphi}^\mrm{H}\hat\omega\(\omega_\mrm{H}\hat\phi-\hat\omega\phi_\mrm{H}\)}{\sqrt{{\mca N}^2_\mrm{H}-g_{\varphi\varphi}^\mrm{H}{\hat\omega}^2}}\rvt\lbl{qbE}}
Looking at
the 
$\left|\lambda\ri|\to\infty$ behavior of Eq.~\eqref{qparam}, one can deduce that
\rov{\hat\phi+\hat\omega A_\varphi^\mrm{H}<0}{qminreq}
corresponds to Eq.~\eqref{qmin} being a lower bound, whereas if the opposite inequality is satisfied, Eq.~\eqref{qmin} will be an upper bound.

\noindent
Case {\bf 1c}: If the expression under the square root in Eq.~\eqref{lamqmin} is zero (and Eq.~\eqref{lamqmin} is thus an invalid expression), the values of charge in the admissible region will be bounded by $q_\mrm{b}=0$. 
However, $q=0$ cannot be attained for any finite value of other parameters on the border.

Let us turn to the angular momentum $L$.
From Eq.~\eqref{Lparam}, we find that a value of $\lambda$ for a candidate stationary point is 
\rov{\lambda=m\frac{g_{\varphi\varphi}^\mrm{H}\hat\phi\sgn A_\varphi^\mrm{H}}{\sqrt{{\mca N}^2_\mrm{H}\(A_\varphi^\mrm{H}\)^2-g_{\varphi\varphi}^\mrm{H}{\hat\phi}^2}}\rvt}{lamLmin}
Again, there are three possible cases.

\noindent
Case {\bf 2a}: If Eq.~\eqref{lamLmin} is imaginary, there is no bound on $L$ in the admissible region.

\noindent
Case {\bf 2b}: If Eq.~\eqref{lamLmin} is real, 
it will correspond to an extremum of Eq.~\eqref{Lparam} with value
\rov{L_\mrm{b}=-\frac{m\sgn A_\varphi^\mrm{H}}{\hat\phi+\hat\omega A_\varphi^\mrm{H}}\sqrt{{\mca N}^2_\mrm{H}\(A_\varphi^\mrm{H}\)^2-g_{\varphi\varphi}^\mrm{H}{\hat\phi}^2}\rvc}{Lmin}
which serves as a bound for $L$.
For Eqs.~\eqref{qparam} and \eqref{Eparam}, the following values of $q$ and $E$ will be implied by Eq.~\eqref{lamLmin}:
\prov{q&=-\frac{m\sgn A_\varphi^\mrm{H}}{\hat\phi+\hat\omega A_\varphi^\mrm{H}}\frac{{\mca N}^2_\mrm{H}A_\varphi^\mrm{H}+g_{\varphi\varphi}^\mrm{H}\hat\omega\hat\phi}{\sqrt{{\mca N}^2_\mrm{H}\(A_\varphi^\mrm{H}\)^2-g_{\varphi\varphi}^\mrm{H}{\hat\phi}^2}}\rvc\lbl{Lbq}\\
E&=\frac{m\sgn A_\varphi^\mrm{H}}{\hat\phi+\hat\omega A_\varphi^\mrm{H}}\frac{{\mca N}^2_\mrm{H}A_t^\mrm{H}A_\varphi^\mrm{H}+g_{\varphi\varphi}^\mrm{H}\hat\phi\(\omega_\mrm{H}\hat\phi-\hat\omega\phi_\mrm{H}\)}{\sqrt{{\mca N}^2_\mrm{H}\(A_\varphi^\mrm{H}\)^2-g_{\varphi\varphi}^\mrm{H}{\hat\phi}^2}}\rvt\lbl{LbE}}
From the $\left|\lambda\ri|\to\infty$ behavior of Eq.~\eqref{Lparam}, we can infer that Eq.~\eqref{Lmin} is a lower bound, if 
\rov{\frac{A_\varphi^\mrm{H}}{\hat\phi+\hat\omega A_\varphi^\mrm{H}}<0\rvt}{Lminreq}
When the opposite inequality is satisfied, Eq.~\eqref{Lmin} is an upper bound.

\noindent
Case {\bf 2c}: If Eq.~\eqref{lamLmin} is undefined due to the expression in the denominator being zero, the values of $L$ in the admissible region will be bounded by $L_\mrm{b}=0$, and this value cannot be reached for a finite value of other parameters on the border.

Combining the possibilities together, we can conclude that cases {\bf 1a2b} and {\bf 1a2c} correspond to the centrifugal mechanism, whereas variants {\bf 1b2a} and {\bf 1c2a} correspond to the electrostatic mechanism.  Case {\bf 1a2a} signifies the coexistence of both. Note that the combination of signs of $q$ and $L$
leading to $\chi<0$ is excluded in any case. The other possible
 combinations, i.e., {\bf 1b2b} and {\bf 1c2c}, 
do not correspond to any simpler limiting cases.

Let us turn to the energy $E$ to finish the discussion of bounds on parameters.
From Eq.~\eqref{Eparam}, we find that a value of $\lambda$ for a candidate stationary point is 
\rov{\lambda=-m\frac{g_{\varphi\varphi}^\mrm{H}\(\omega_\mrm{H}\hat\phi-\hat\omega\phi_\mrm{H}\)\sgn A_t^\mrm{H}}{\sqrt{{\mca N}^2_\mrm{H}\(A_t^\mrm{H}\)^2-g_{\varphi\varphi}^\mrm{H}\(\omega_\mrm{H}\hat\phi-\hat\omega\phi_\mrm{H}\)^2}}\rvt}{lamEmin}
Unlike in the previous two cases, this value can be adjusted using the available gauge freedom.
Consequently, we can choose Eq.~\eqref{lamEmin} to be real, as explained below. Furthermore, it turns out that we can also choose the corresponding stationary point of Eq.~\eqref{Eparam} to be a minimum. 
Its value is
\rov{E_\mrm{min}=\frac{m\sgn A_t^\mrm{H}}{\hat\phi+\hat\omega A_\varphi^\mrm{H}}\sqrt{{\mca N}^2_\mrm{H}\(A_t^\mrm{H}\)^2-g_{\varphi\varphi}^\mrm{H}\(\omega_\mrm{H}\hat\phi-\hat\omega\phi_\mrm{H}\)^2}\rvc}{Emin}
and for Eqs.~\eqref{qparam} and \eqref{Lparam},
the following values of $q$ and $L$ are implied by Eq.~\eqref{lamEmin}:
\prov{q=-\frac{m\sgn A_t^\mrm{H}}{\hat\phi+\hat\omega A_\varphi^\mrm{H}}\frac{{\mca N}^2_\mrm{H}A_t^\mrm{H}-g_{\varphi\varphi}^\mrm{H}\hat\omega\(\omega_\mrm{H}\hat\phi-\hat\omega\phi_\mrm{H}\)}{\sqrt{{\mca N}^2_\mrm{H}\(A_t^\mrm{H}\)^2-g_{\varphi\varphi}^\mrm{H}\(\omega_\mrm{H}\hat\phi-\hat\omega\phi_\mrm{H}\)^2}}\rvc\lbl{Eminq}\\L=-\frac{m\sgn A_t^\mrm{H}}{\hat\phi+\hat\omega A_\varphi^\mrm{H}}\frac{{\mca N}^2_\mrm{H}A_t^\mrm{H}A_\varphi^\mrm{H}+g_{\varphi\varphi}^\mrm{H}\hat\phi\(\omega_\mrm{H}\hat\phi-\hat\omega\phi_\mrm{H}\)}{\sqrt{{\mca N}^2_\mrm{H}\(A_t^\mrm{H}\)^2-g_{\varphi\varphi}^\mrm{H}\(\omega_\mrm{H}\hat\phi-\hat\omega\phi_\mrm{H}\)^2}}\rvt\lbl{EminL}}
What are the requirements in order to have a lower bound on $E$ in the admissible region, and is it always possible to make these requirements 
satisfied simultaneously?
First, we have to impose the condition
\rov{{\mca N}^2_\mrm{H}\(A_t^\mrm{H}\)^2-g_{\varphi\varphi}^\mrm{H}\(\omega_\mrm{H}\hat\phi-\hat\omega\phi_\mrm{H}\)^2>0}{ergextg}
to make Eq.~\eqref{lamEmin} real. 
By checking the $\left|\lambda\ri|\to\infty$ behavior of Eq.~\eqref{Eparam}, we can see that 
we must also require
\rov{\frac{A_t^\mrm{H}}{\hat\phi+\hat\omega A_\varphi^\mrm{H}}>0\rvc}{ergming}
in order for Eq.~\eqref{Emin} to be a lower bound.
Next, recalling \eqref{phigen}, one can observe that the combinations $A_t^\mrm{H}\equiv-\phi_\mrm{H}-\omega_\mrm{H}A_\varphi^\mrm{H}$ and $\omega_\mrm{H}\hat\phi-\hat\omega\phi_\mrm{H}$ are linearly independent, 
except for the degenerate case when Eq.~\eqref{lindeg} holds, which has to be treated separately anyway.
Therefore, there is always a way to choose values of $\phi_\mrm{H}$ and $\omega_\mrm{H}$ 
that make any of the conditions
given in Eqs.~\eqref{ergextg} and \eqref{ergming} satisfied or violated.

\subsubsection{Additional remarks}

\lbl{odd:addrem}

Above, we have identified points on the border of the admissible region where a minimal or maximal value of one of the parameters $q$, $L$, and $E$ is reached. 
Values of all the parameters at such points are proportional to the particle's mass. This illustrates the fact that only a reduced set of parameters is needed to describe particles' kinematics. 
In particular, for massive critical particles, two parameters are sufficient. These can be either $\tilde \chi=\frac{\chi}{m}$ and $\tilde\lambda=\frac{\lambda}{m}$ or any two of $\tilde q$,
$l$, and $\varepsilon$. 
Therefore, we can understand the admissible region given in Eq.~\eqref{admreg} as an area in a two-dimensional parameter space. Its border given in Eq.~\eqref{admhypxlam} can be viewed as a curve therein, namely, a branch of a hyperbola with axes $\tilde \chi=0$ and $\tilde \lambda=0$ and with its vertex on $\tilde\lambda=0$. Let us note that in variables $\tilde q$ and $l$ the asymptotes of this hyperbola can be expressed as
\rov{l=-\tilde q\frac{\sqrt{g_{\varphi\varphi}^\mrm{H}}\hat\phi\pm\mca N_\mrm{H}A_\varphi^\mrm{H}}{\sqrt{g_{\varphi\varphi}^\mrm{H}}\hat\omega\mp\mca N_\mrm{H}}\rvs}{admhypaskn}
see Eq.~(43) in \cite{a2}.

Considering the tilde parameters, i.e., the parameters
normalized to unit rest mass, one excludes \emph{a priori}
critical photons. 
However, this is not a big issue, since they have trivial kinematics. Indeed, all critical photons 
 share the same single value of impact parameter $b_\mrm{cr}=\frac{1}{\omega_\mrm{H}}$. Therefore, 
the parameter space of critical photons is effectively zero dimensional, and their kinematics depends only on the properties of the spacetime itself.
When are the critical photons able to approach $r_\mrm{H}$? The expansion
given in Eq.~\eqref{znhcr} reads for those critical photons
\rov{\mca Z\approx\left|L\ri|\sqrt{{\hat\omega}^2-\frac{{\mca N}^2_\mrm{H}
}{g_{\varphi\varphi}^\mrm{H}}}\(r-r_\mrm{H}\)+\dots}{}
Here, the expression under the square root is proportional to the one in Eq.~\eqref{lamqmin} with a negative factor. Therefore, critical photons can be involved in the high-energy collisional processes close to $r_\mrm{H}$ only in the case {\bf 1a}.

Finally, let us clarify the link between bounds on $q, L, E$ and restrictions on signs of those parameters. Starting with $q$, we can observe that condition
\eqref{qminreq} also determines the sign of Eq.~\eqref{qmin}. Thus, 
if Eq.~\eqref{qmin} is a lower bound, its value is positive, whereas if it is an upper bound, its value is negative. Therefore, 
whenever values of $q$ in the admissible region are bounded by Eq.~\eqref{qmin}, 
they 
must also all have 
the same sign. 
An identical relation holds 
between Eq.~\eqref{Lmin} and \eqref{Lminreq}. Last, the gauge condition \eqref{ergming}, 
which we use to enforce a lower bound on energy, also implies that the bound given in
Eq.~\eqref{Emin} 
has a positive value. 
Hence, dividing by $m$ and rearranging the sign factors, we can express the (possible) bounds on values of
$\tilde q$, $l$, and $\varepsilon$ in the admissible region as follows:
\prov{-\tilde q\sgn\f(\hat\phi+\hat\omega A_\varphi^\mrm{H}\)&>\frac{1}{\left|\hat\phi+\hat\omega A_\varphi^\mrm{H}\ri|}\sqrt{
{\mca N}^2_\mrm{H}
-g_{\varphi\varphi}^\mrm{H}{\hat\omega}^2}\rvc\lbl{tqmin}\\
-l\sgn\f[\(\hat\phi+\hat\omega A_\varphi^\mrm{H}\)A_\varphi^\mrm{H}\]&>\frac{1}{\left|\hat\phi+\hat\omega A_\varphi^\mrm{H}\ri|}\sqrt{{\mca N}^2_\mrm{H}\(A_\varphi^\mrm{H}\)^2-g_{\varphi\varphi}^\mrm{H}{\hat\phi}^2}\rvc\lbl{lmin}\\
\varepsilon&>\frac{1}{\left|\hat\phi+\hat\omega A_\varphi^\mrm{H}\ri|}\sqrt{{\mca N}^2_\mrm{H}\(A_t^\mrm{H}\)^2-g_{\varphi\varphi}^\mrm{H}\(\omega_\mrm{H}\hat\phi-\hat\omega\phi_\mrm{H}\)^2}\rvt\lbl{epsmin}}{}

\subsection{Degenerate case}

\lbl{odd:admdeg}

Let us now explore the previously 
 excluded case when the degeneracy condition \eqref{lindeg} is satisfied.
As we noted 
in Sec.~\ref{odd:nhcr}, 
this means that the variables $\chi$ and $\lambda$ 
become proportional, 
namely, $\chi=-\hat\omega\lambda$. 
Because of this, Eq.~\eqref{admhypxlam} with Eq.~\eqref{lindeg} degenerates into an algebraic equation for one 
variable, 
which has a single solution, 
\rov{\lambda=-\frac{m\sgn\hat\omega}{\sqrt{\frac{{\hat\omega}^2}{{\mca N}^2_\mrm{H}}-\frac{1}{g_{\varphi\varphi}^\mrm{H}}}}\rvt}{admlinlam}
One can see that the expressions under the square roots in the denominators of
Eq.~\eqref{lamqmin} and
Eq.~\eqref{admlinlam} are related by a negative factor. Therefore, Eq.~\eqref{admlinlam} 
is defined in real numbers in case {\bf 1a}.
On the other hand, in cases {\bf 1b} and {\bf 1c}, there is no real solution of Eq.~\eqref{admhypxlam} 
with Eq.~\eqref{lindeg}, and thus 
the collisional processes studied here are impossible for critical particles with any value of $\lambda$.
We are unaware of a black hole spacetime 
 where this would occur in the equatorial plane. However, a similar thing happens around the poles of the Kerr solution, as demonstrated by \cite{HK11b}.
 Note also that 
$\lambda\sgn\hat\omega>0$ certainly violates 
Eq.~\eqref{admreg}
 with Eq.~\eqref{lindeg}.

Let us now consider the physical interpretation of
the degenerate case given by
Eq.~\eqref{admlinlam}. Using Eq.~\eqref{nhparinvdeg} (with $\mca X_\mrm{H}=0$),
we find that 
 Eq.~\eqref{admlinlam}
can be expressed as \prov{L&=-\frac{m\sgn\hat\omega}{\sqrt{\frac{{\hat\omega}^2}{{\mca N}^2_\mrm{H}}-\frac{1}{g_{\varphi\varphi}^\mrm{H}}}}+qA_\varphi^\mrm{H}\rvc\lbl{admlinL}\\
E&=-\frac{m\omega_\mrm{H}\sgn\hat\omega}{\sqrt{\frac{{\hat\omega}^2}{{\mca N}^2_\mrm{H}}-\frac{1}{g_{\varphi\varphi}^\mrm{H}}}}-qA_t^\mrm{H}\rvt\lbl{admlinE}}
The charge of the particle 
plays a role of a free 
variable; 
 there can never be a bound on the values of $q$ in the admissible region in the degenerate case. 
However, if $A_\varphi^\mrm{H}=0$, 
Eq.~\eqref{admlinL} will correspond to a single value of $L$, 
which will constitute a
bound on $L$. If Eq.~\eqref{lindeg} holds together with $A_\varphi^\mrm{H}=0$, we have $-\sgn\f(\hat\omega L\)=\sgn \chi>0$, 
and therefore 
we can infer
\rov{-l\sgn\hat\omega>\frac{1}{\sqrt{\frac{{\hat\omega}^2}{{\mca N}^2_\mrm{H}}-\frac{1}{g_{\varphi\varphi}^\mrm{H}}}}\rvt}{lmindeg}
Note that Eq.~\eqref{lindeg} with $A_\varphi^\mrm{H}=0$ implies $\hat\phi=0$. This would signify case {\bf 2c} as defined in general (cf. \cite{a2}), yet the bound on $l$ has a nonzero value in this case.
We want a lower bound on $E$ in the admissible region, and therefore we impose gauge conditions $A_t^\mrm{H}=0$ and $\omega_\mrm{H}\sgn\hat\omega<0$ in Eq.~\eqref{admlinE}.
Then, it holds that
\rov{\varepsilon>\frac{\left|\omega_\mrm{H}\ri|}{{\sqrt{\frac{{\hat\omega}^2}{{\mca N}^2_\mrm{H}}-\frac{1}{g_{\varphi\varphi}^\mrm{H}}}}}\rvt}{epsmindeg}

\section{Energy extraction}

\lbl{sek:extra}

\subsection{Conservation laws and kinematic regimes}

\subsubsection{Conservation laws}

Now, we discuss the properties of particles than can be produced in the
high-energy collisional processes described in Sec.~\ref{odd:BSW} and,
in particular, how much energy such particles can extract from a black
hole.
Let us consider a simple setup in which a
critical particle or  a
nearly critical particle, call it particle $1$, 
and an incoming usual particle, call it particle $2$,
collide close to the horizon radius $r_\mrm{H}$, 
and their interaction leads to production of just
two new particles, particle $3$ and particle $4$.
We
assume conservation of charge
\rov{q_1+q_2=q_3+q_4\rvc}{}
and also the conservation of
4-momentum at the point of collision. From
the azimuthal component
of the 4-momentum, see Eq.~\eqref{constm}, we infer the conservation of angular momentum
\rov{L_1+L_2=L_3+L_4\rvt}{}
The conservation law for the time component of the 4-momentum can be used to derive
the conservation of energy
\rov{E_1+E_2=E_3+E_4\rvt}{}
There is another conserved 4-momentum component, the radial one $p^r$.  Instead of writing down the conservation of $p^r$, we combine it together with the conservation of the time component $p^t$ of the 4-momentum.
For that, we use the combinations $N^2p^t\mp N\sqrt{g_{rr}}p^r$, which
is advantageous because they lead to combinations of the functions $\mca X$ and $\mca Z$, both defined
in Eq.~\eqref{aux1}. Indeed, 
\rov{N^2p^t\mp N\sqrt{g_{rr}}p^r=\mca X\mp\sigma \mca Z\rvt}{}
Since we assumed that particle $2$ is incoming, $\sigma_2=-1$, 
the 
summation of the conservation laws leads to the following equation:
\rov{\mca X_1\mp\sigma_1\mca Z_1+\mca X_2\pm\mca Z_2=\mca X_3\mp\sigma_3\mca Z_3+\mca X_4\mp\sigma_4\mca Z_4\rvt}{trsum}
Let us find the leading-order terms in Eq.~\eqref{trsum}.
For usual particles, $\mca X$ and $\mca Z$ differ by a term proportional to $N^2$ (see Eq.~\eqref{znhus}), so their combinations with different signs have different leading orders in expansion around $r_\mrm{H}$,
\prov{\mca X-\mca Z&\sim \(r-r_\mrm{H}\)^2\rvc &\mca X+\mca Z\approx2\mca X_\mrm{H}\rvt}
On the other hand, in the case of
critical particles, or nearly critical particles,
the leading order of expansion in $r_\mrm{C}-r_\mrm{H}$ for both combinations is 
\rov{\mca X\mp\mca Z\sim \(r_\mrm{C}-r_\mrm{H}\)\rvt}{}
Let us start to analyze Eq.~\eqref{trsum} with its upper sign. 
We assumed particle $2$ to be a
usual particle, and thus the leading order is the zeroth one, so 
\rov{2\mca X_2^\mrm{H}=\mca X_3^\mrm{H}-\sigma_3\mca Z_3^\mrm{H}+\mca X_4^\mrm{H}-\sigma_4\mca Z_4^\mrm{H}\rvt}{cons1}
This equation can be satisfied only when one of the final particles, say $4$, is usual and incoming, i.e., $\mca X_4^\mrm{H}>0$, $\sigma_4=-1$.
Let us now turn to  Eq.~\eqref{trsum} with its lower sign.
We see that usual incoming particles $2$ and $4$ will make no contribution to zeroth order and first order. 
On the other hand, critical particle $1$ will contribute to the first order, and this contribution will dominate the left-hand side. Therefore, the 
expansion of the right-hand side must also be dominated by a first-order contribution, which means that particle $3$ has to be critical or nearly critical. 
The leading order of Eq.~\eqref{trsum} with the lower  sign thus becomes 
\rov{\chi_1+\sigma_1\sqrt{\chi_1^2-{\mca N}^2_\mrm{H}\(m_1^2+\frac{\lambda_1^2}{g_{\varphi\varphi}^\mrm{H}}\)}=\chi_3-C_3+\sigma_3\sqrt{\(\chi_3-C_3\)^2-{\mca N}^2_\mrm{H}\(m_3^2+\frac{\lambda_3^2}{g_{\varphi\varphi}^\mrm{H}}\)}\rvt}{consfin}
Here, $C_3$ 
parametrizes deviation of particle $3$ from criticality according to Eq.~\eqref{fwhornc}. We put $C_1=0$ for simplicity. 
One can denote the whole left-hand side of Eq.~\eqref{consfin} 
as a new parameter $\msi A_1$ such that
\rov{\mca N_\mrm{H}\msi A_1\equiv \chi_1+\sigma_1\sqrt{\chi_1^2-{\mca N}^2_\mrm{H}\(m_1^2+\frac{\lambda_1^2}{g_{\varphi\varphi}^\mrm{H}}\)}\rvc}{A1}
which will 
carry all the 
information about particle $1$. Since $\chi_1>0$, we can make sure that $\msi A_1\geqslant0$. The difference between BSW-type processes,
$\sigma_1=-1$, and Schnittman-type processes, $\sigma_1=+1$, is absorbed into the definition of $\msi A_1$, and thus 
the results expressed using $\msi A_1$ hereafter will be the same for both.

\subsubsection{Kinematic regimes}

Having derived Eqs.~\eqref{cons1} and \eqref{consfin}
from the conservation law, Eq.~\eqref{trsum}, we now turn to
their physical implications
in collisional Penrose processes.
For a Penrose process, one of the particles must fall inside the black
hole, and we can make sure that particle $4$ is bound to do so
according to Eq.~\eqref{cons1}.
On the other hand, particle 3 can be
produced in four distinct kinematic regimes, based on the combination
of sign of $C_3$ and the sign variable $\sigma_3$
in Eq.~\eqref{consfin}.  In accordance with
\cite{Zasl12c}, let us denote the regimes with $C_3>0$ as $+$,
$C_3<0$ as $-$, $\sigma_3=+1$ as out, and $\sigma_3=-1$ as
in.
The four kinematic regimes are then
out+,
out$-$,
in+, and 
in$-$.

There are important differences among the four kinematic regimes in several regards. First, we should determine which ones allow particle $3$ to escape from the vicinity of the black hole.
For simplicity, let us assume a situation when condition \eqref{mallow} is well approximated by linear expansion terms. 
In such a case, there can be at most one turning point near 
$r_\mrm{H}$. The radius $r_\mrm{T}$ of this turning point is defined by the condition
\rov{\chi_3\(r_\mrm{T}-r_\mrm{H}\)-C_3\(r_\mrm{C}-r_\mrm{H}\)={\mca N}_\mrm{H}\sqrt{m_3^2+\frac{\lambda_3^2}{g_{\varphi\varphi}^\mrm{H}}}\(r_\mrm{T}-r_\mrm{H}\)\rvc}{}
which can be rearranged as follows:
\rov{r_\mrm{C}-r_\mrm{T}=\(r_\mrm{C}-r_\mrm{H}\)\frac{\chi_3-C_3-{\mca N}_\mrm{H}\sqrt{m_3^2+\frac{\lambda_3^2}{g_{\varphi\varphi}^\mrm{H}}}}{\chi_3-{\mca N}_\mrm{H}\sqrt{m_3^2+\frac{\lambda_3^2}{g_{\varphi\varphi}^\mrm{H}}}}\rvt}{nhturn}
Note that 
Eq.~\eqref{nhturn} may imply $r_\mrm{T}<r_\mrm{H}$, and hence no turning point in the region of our interest.
The motion of particle $3$ must be allowed at $r_\mrm{C}$, where it is produced; hence,
\rov{\chi_3-C_3-{\mca N}_\mrm{H}\sqrt{m_3^2+\frac{\lambda_3^2}{g_{\varphi\varphi}^\mrm{H}}}>0\rvt}{}
Therefore, the numerator of the fraction in Eq.~\eqref{nhturn} is positive,
and since $r_\mrm{C}>r_\mrm{H}$ by definition, we can conclude that $r_\mrm{T}<r_\mrm{C}$ for particles produced with parameters in the admissible region, whereas $r_\mrm{T}>r_\mrm{C}$ for the ones outside of it. (Note the definition of the admissible region of parameters given in Eq.~\eqref{admreg}.) However, if $r_\mrm{T}>r_\mrm{C}$, particle $3$ produced at $r_\mrm{C}$ 
can never escape.
Therefore, in order for particle $3$ to escape, it must be produced with parameters in the admissible region.

Particles with $C_3>0$ cannot fall into the black hole by definition, and thus they must have a turning point at a radius $r_\mrm{H}<r_\mrm{T}<r_\mrm{C}$. Therefore, in regimes out$+$ and in$+$, particle 
$3$ can be produced only with parameters in the admissible region, and 
it is automatically guaranteed to escape. 

Particles with $C_3<0$, in turn, can cross the horizon; their motion is allowed both at $r_\mrm{H}$ and at $r_\mrm{C}$. Hence, there must be an even number of turning points between $r_\mrm{H}$ and $r_\mrm{C}$. However, we assumed the existence of at most one turning point, and thus there can be none. Incoming particle $3$ produced with $C_3<0$ therefore has to fall into the black hole; i.e., escape in the in$-$ regime is impossible.
Last, in the out$-$ regime, particle $3$ can either escape or be reflected and fall into the black hole, based on whether its parameters lie in the admissible region or not.

The way in which parameters $C_3$ and $\sigma_3$ determine escape
possibilities of particle $3$ is actually independent of the
particular system in question. This can be seen, e.g., through
comparison with Sec.~IV B in \cite{a3}, in which particles moving along
the symmetry axis are considered.  However, despite being so universal
and so important for escape of particle $3$, parameters $C_3$ and
$\sigma_3$ are quite irrelevant for all other purposes.  Indeed, if
particle $3$ escapes, $\sigma_3$ must eventually flip to $+1$, whereas
$C_3$ encodes only a small deviation from fine tuning of parameters of
particle $3$.

We shall now solve Eq.~\eqref{consfin} for $C_3$ and $\sigma_3$, 
in order to 
view the four different kinematic regimes in terms of the other parameters,
i.e.,  $\chi_3$, $\lambda_3$, $m_3$, 
and $\msi A_1$. 
First, we can observe from Eq.~\eqref{consfin} that 
\rov{\sigma_3=\sgn\f(\mca N_\mrm{H}\msi A_1-\chi_3+C_3\)\rvt}{sig3_0}
Expressing $C_3$ form Eq.~\eqref{consfin} and then substituting 
it back into Eq.~\eqref{sig3_0}, we obtain the solutions as follows:
\prov{C_3&=\chi_3-\frac{{\mca N}_\mrm{H}}{2}\[\msi A_1+\frac{1}{\msi A_1}\(m_3^2+\frac{\lambda_3^2}{g_{\varphi\varphi}^\mrm{H}}\)\]\lbl{c3}\rvc\\
\sigma_3&=\sgn\f[\msi A_1^2-\(m_3^2+\frac{\lambda_3^2}{g_{\varphi\varphi}^\mrm{H}}\)\]\lbl{sig3}\rvt}
Since we are interested only in the sign of $C_3$, and $\sigma_3$ is a sign variable \emph{per se}, only ratios among the four parameters on the right-hand sides matter to us. Therefore, 
 we have considerable freedom in 
choosing 
the relevant three variables. 
Nevertheless, we have seen above that we also need to 
consider the admissible region, 
for which the relevant 
parameters are $\tilde\chi,\tilde\lambda$. Thus, 
it is natural to understand Eqs.~\eqref{c3} and \eqref{sig3} as depending on $\tilde\chi_3, \tilde\lambda_3$ and on the ratio between $\msi A_1$ and $m_3$.

A third parameter, i.e., the ratio between $\msi A_1$ and $m_3$, 
clearly stands out; it 
tracks a comparison between properties of two particles, and it is irrelevant for the admissible region of particle $3$. Therefore, we find it natural to visualize the different kinematic regimes as regions in the same  
two-dimensional parameter space as the admissible region, with the ratio between $\msi A_1$ and $m_3$ serving as an external parameter.
However, 
since we are interested in a
physical interpretation, namely, in energy extraction, we will keep $\msi A_1$ and $m_3$ separate in the equations, and we will not explicitly pass to the parameters normalized to unit rest mass.

If we treat the ratio between $\msi A_1$ and $m_3$ as an external parameter, 
there are just two main possibilities, namely, a heavy regime and a light regime.
In the heavy regime, defined by 
$m_3>\msi A_1$, the right-hand side of Eq.~\eqref{sig3} is negative for any $\tilde\lambda_3$, and hence the in region covers the whole parameter space. 
In the light regime, defined by $m_3<\msi A_1$, the parameter space is divided into in and out regions.

\subsection{Structure of the parameter space}

\lbl{odd:exstps}

\subsubsection{Overall picture}

Now, we should understand how the regions of parameters corresponding to different kinematic regimes are distributed across 
our parameter space. 
 Let us start with 
the distinction between {$+$} and {$-$} regimes, which is always present
regardless of the ratio between $\msi A_1$ and $m_3$. 
From the solution given in Eq.~\eqref{c3}, 
we see that $C_3>0$ implies the inequality
\rov{\chi_3>\frac{{\mca N}_\mrm{H}}{2}\[\msi A_1+\frac{1}{\msi A_1}\(m_3^2+\frac{\lambda_3^2}{g_{\varphi\varphi}^\mrm{H}}\)\]\rvc}{plusreg}
which defines the $+$ region of parameters.
Conversely, the inequality opposite to Eq.~\eqref{plusreg} 
entails $C_3<0$ and defines the $-$ region. 
For $C_3=0$, one has
\rov{\chi_3=\frac{{\mca N}_\mrm{H}}{2}\[\msi A_1+\frac{1}{\msi A_1}\(m_3^2+\frac{\lambda_3^2}{g_{\varphi\varphi}^\mrm{H}}\)\]\rvc}{C3zeroxlam}
which defines the border between the regions, and it 
corresponds to particle $3$ being produced as precisely critical. 
In the $\tilde\chi_3, \tilde\lambda_3$ parameter space, Eq.~\eqref{C3zeroxlam} represents a parabola 
with axis $\tilde\lambda_3=0$. 

For a physical interpretation, let us substitute Eq.~\eqref{C3zeroxlam} into Eq.~\eqref{nhparinv} to obtain parametric expressions for the border as follows:
\prov{q_3&=-\frac{1}{\hat\phi+\hat\omega A_\varphi^\mrm{H}}\left\{\hat\omega\lambda_3+\frac{{\mca N}_\mrm{H}}{2}\[\msi A_1+\frac{1}{\msi A_1}\(m_3^2+\frac{\lambda_3^2}{g_{\varphi\varphi}^\mrm{H}}\)\]\ri\}\lbl{C3zeroq}\rvc\\
L_3&=\frac{1}{\hat\phi+\hat\omega A_\varphi^\mrm{H}}\left\{\hat\phi\lambda_3-\frac{{\mca N}_\mrm{H}A_\varphi^\mrm{H}}{2}\[\msi A_1+\frac{1}{\msi A_1}\(m_3^2+\frac{\lambda_3^2}{g_{\varphi\varphi}^\mrm{H}}\)\]\ri\}\lbl{C3zeroL}\rvc\\
E_3&=\frac{1}{\hat\phi+\hat\omega A_\varphi^\mrm{H}}\left\{\(\omega_\mrm{H}\hat\phi-\hat\omega\phi_\mrm{H}\)\lambda_3+\frac{{\mca N}_\mrm{H}A_t^\mrm{H}}{2}\[\msi A_1+\frac{1}{\msi A_1}\(m_3^2+\frac{\lambda_3^2}{g_{\varphi\varphi}^\mrm{H}}\)\]\ri\}\rvt\lbl{C3zeroE}}
Recalling the gauge condition Eq.~\eqref{ergming}, we can make sure that Eq.~\eqref{C3zeroE} leads to $E_3\to\infty$ for $\left|\lambda_3\ri|\to\infty$. 
Therefore, we see that 
values of $E_3$ 
in neither the {$+$} 
nor {$-$} region 
are bounded from above. This was not
possible in the previously
known special cases; see \cite{HaNeMi}.
Since the escape of particle $3$ is guaranteed in the {$+$} regime, we can also 
conclude that there is no upper bound on the energy extracted from the black hole. 
Such a possibility is often called the super-Penrose process.
Furthermore, as the {$+$} region exists for any value of $m_3$, we see that there is no bound on the mass of escaping particles
as well. 

Now, let us turn to the distinction between in and out regimes. From Eq.~\eqref{sig3}, we can see that parameters in the in region must satisfy the condition
\rov{\left|\lambda_3\ri|>\sqrt{g_{\varphi\varphi}^\mrm{H}\({\msi A_1^2}-m_3^2\)}\rvc}{inreg}
whereas the opposite inequality holds for parameters in the out region.
The two values of $\lambda_3$ that separate the regions, i.e.,
\rov{\lambda_3=\pm\sqrt{g_{\varphi\varphi}^\mrm{H}\({\msi A_1^2}-m_3^2\)}\rvc}{sig3zero}
correspond to a situation when our leading-order approximation breaks down, since the square root on the right-hand side of Eq.~\eqref{consfin} becomes zero and 
we cannot consistently assign a value to $\sigma_3$. This indicates that  particle $3$ with those values of $\lambda_3$ will be produced as class II nearly critical, 
and a different expansion would be needed to determine its initial direction of motion.

Let us note that particle $3$ can be produced in the in$+$ regime,
i.e., with $\chi_3$ and $\lambda_3$ satisfying both
Eq.~\eqref{plusreg} and Eq.~\eqref{inreg},
for any value of the ratio between $m_3$ and $\msi A_1$. This is another thing that was not possible in the previously studied special cases (i.e., vacuum black holes and nonrotating black holes).

\subsubsection{Osculation points}

\enlargethispage{-\baselineskip}

Having derived borders that divide the $\tilde\chi_3,\tilde\lambda_3$ parameter space according to various criteria, we shall now 
consider the {corners} where the borders meet. 
We can get insight into this issue from the physical interpretation of the borders;
Eq.~\eqref{admhypxlam} 
gives a set of parameters for which precisely critical particles are of class II,
 Eq.~\eqref{C3zeroxlam} corresponds to particle $3$ being produced as precisely critical, and Eq.~\eqref{sig3zero} corresponds to particle $3$ being produced as class II critical or nearly critical. 
  If any two of those 
 eventualities happen together, 
  the third one follows automatically. 
  Therefore, all three borders must meet in the same points of the parameter space. Indeed, substituting Eq.~\eqref{sig3zero} into both Eq.~\eqref{admhypxlam} and Eq.~\eqref{C3zeroxlam} leads to $\chi_3=\mca N_\mrm{H}\msi A_1$. 
  Conversely, in the {heavy regime} $m_3>\msi A_1$, in which case Eq.~\eqref{sig3zero} is absent, the 
  remaining borders given in Eqs.~\eqref{admhypxlam} and \eqref{C3zeroxlam} 
  cannot meet at any point. Note that in the $m_3=\msi A_1$ case, Eqs.~\eqref{admhypxlam} and \eqref{C3zeroxlam} touch at $\lambda_3=0$.

We have also seen that particle $3$ can be produced with $C_3>0$ only when 
its other parameters satisfy the condition \eqref{admreg}. Therefore, the {$+$} region must lie inside the admissible region in the parameter space, and their borders can only osculate. 
One can make sure that 
this is indeed the case by comparing the limiting behavior of Eqs.~\eqref{admhypxlam} and \eqref{C3zeroxlam} for $\left|\lambda_3\ri|\to\infty$ and their values at $\lambda_3=0$, i.e., in between the values given in Eq.~\eqref{sig3zero}.
By putting Eq.~\eqref{sig3zero} into Eqs.~\eqref{C3zeroq}-\eqref{C3zeroE} (or into Eqs.~\eqref{qparam}-\eqref{Eparam}), we obtain
the values of $q$, $L$, and $E$ for the osculating points, namely,
\prov{q&=\frac{1}{\hat\phi+\hat\omega A_\varphi^\mrm{H}}\[\mp\hat\omega\sqrt{g_{\varphi\varphi}^\mrm{H}\({\msi A_1^2}-m_3^2\)}-{\mca N}_\mrm{H}\msi A_1\]\rvc\lbl{oscq}\\
L&=\frac{1}{\hat\phi+\hat\omega A_\varphi^\mrm{H}}\[\pm\hat\phi\sqrt{g_{\varphi\varphi}^\mrm{H}\({\msi A_1^2}-m_3^2\)}-{\mca N}_\mrm{H}A_\varphi^\mrm{H}\msi A_1\]\rvc\\
E&=\frac{1}{\hat\phi+\hat\omega A_\varphi^\mrm{H}}\[\pm\(\omega_\mrm{H}\hat\phi-\hat\omega\phi_\mrm{H}\)\sqrt{g_{\varphi\varphi}^\mrm{H}\({\msi A_1^2}-m_3^2\)}+{\mca N}_\mrm{H}A_t^\mrm{H}\msi A_1\]\rvt\lbl{oscE}}

\subsubsection{Bounds on parameters: General considerations}

\lbl{odd:exbp}

We have seen that there is no upper bound on the energy $E_3$ in the regions of the parameter space which correspond to particle $3$ being able to escape. Let us now search for other bounds on the parameters of particle $3$ in these regions.
There are multiple possibilities, depending on the ratio between $\msi A_1$ and $m_3$.

First, let us consider a hypothetical interaction, for which this ratio can take any value. 
More precisely, we shall consider an idealized scenario, in which it is possible to produce particle $3$ with any value of $m_3$ in the processes with the same fixed value of $\msi A_1$.
 (Note that keeping $\msi A_1$ fixed is motivated by existence of upper bounds on $\msi A_1$ in
 terms of $E_1$; see Eqs.~\eqref{A1ucbBSW}
and \eqref{A1ucbSch}. Moreover, one can also find a lower bound on $\msi A_1$ in a similar manner for $\sigma_1=+1$.)
 Now, let us look at the union of all the {$+$} regions corresponding to all the possible values of $m_3$. Since the osculation points given by Eq.~\eqref{sig3zero} can occur at any value of $\tilde\lambda_3$, 
 we see that this union will fill the whole admissible region in $\tilde\chi_3,\tilde\lambda_3$ space. Therefore, the possible bounds
 given in Eqs.~\eqref{tqmin}-\eqref{epsmin} on $\tilde q$, $l$, and $\varepsilon$ in the admissible region will also 
 serve as bounds on $\tilde q_3$, $l_3$, and $\varepsilon_3$ of particles 
  produced by our hypothetical interaction. 

Second, 
let us consider a more realistic scenario, in which only some values of the ratio between $m_3$ and $\msi A_1$ are possible. In such a case, we can 
search for bounds on parameters in the {$+$} region for given values of $m_3$ and $\msi A_1$.
Since the parametric expressions given in Eqs.~\eqref{C3zeroq}-\eqref{C3zeroE} are mere quadratic functions of $\lambda_3$, they will always reach an extremum, and therefore there will always be bounds on values of
$q_3$,
$L_3$, and $E_3$ in the {$+$} region.
Starting with $q_3$, we find that for 
\rov{\lambda_3=-\frac{g_{\varphi\varphi}^\mrm{H}\hat\omega}{{\mca N}_\mrm{H}}\msi A_1\rvc}{q3min}
Eq.~\eqref{C3zeroq} reaches an extremum with value
\rov{q_3^\mrm{b}=-\frac{1}{2\(\hat\phi+\hat\omega A_\varphi^\mrm{H}\)}\[\mca N_\mrm{H}\(\msi A_1+\frac{m_3^2}{\msi A_1}\)-\frac{g_{\varphi\varphi}^\mrm{H}{\hat\omega}^2\msi A_1}{{\mca N}_\mrm{H}}\]\rvt}{}
Turning to $L_3$, we can infer that for 
\rov{\lambda_3=\frac{g_{\varphi\varphi}^\mrm{H}\hat\phi}{{\mca N}_\mrm{H}A_\varphi^\mrm{H}}\msi A_1\rvc}{}
expression \eqref{C3zeroL} reaches an extremum with value
\rov{L_3^\mrm{b}=-\frac{1}{2\(\hat\phi+\hat\omega A_\varphi^\mrm{H}\)}\[\mca N_\mrm{H}A_\varphi^\mrm{H}\(\msi A_1+\frac{m_3^2}{\msi A_1}\)-\frac{g_{\varphi\varphi}^\mrm{H}{\hat\phi}^2\msi A_1}{{\mca N}_\mrm{H}A_\varphi^\mrm{H}}\]\rvt}{L3min}
For $E_3$, the situation is again different due to dependence on gauge. Looking at the $\left|\lambda_3\ri|\to\infty$ behavior of Eq.~\eqref{C3zeroE}, we can make sure that the condition \eqref{ergming} implies that Eq.~\eqref{C3zeroE} will reach a minimum. It occurs for
\rov{\lambda_3=-\frac{g_{\varphi\varphi}^\mrm{H}\msi A_1}{{\mca N}_\mrm{H}A_t^\mrm{H}}\(\omega_\mrm{H}\hat\phi-\hat\omega\phi_\mrm{H}\)\rvc}{}
and its value is
\rov{E_3^\mrm{min}=\frac{1}{2\(\hat\phi+\hat\omega A_\varphi^\mrm{H}\)}\[\mca N_\mrm{H}A_t^\mrm{H}\(\msi A_1+\frac{m_3^2}{\msi A_1}\)-\frac{g_{\varphi\varphi}^\mrm{H}\msi A_1}{{\mca N}_\mrm{H}A_t^\mrm{H}}\(\omega_\mrm{H}\hat\phi-\hat\omega\phi_\mrm{H}\)^2\]\rvt}{E3min}

\subsubsection{Additional remarks on the out$-$ region}

\enlargethispage{-\baselineskip}

The discussion above can be extended by 
analyzing bounds on parameters in 
further, special regions in the parameter space. The out$-$ region 
 is particularly interesting in this regard, since there is an upper bound on the values of energy $E_3$ in this region. As noted in \cite{a3}, this can be used to illustrate the difference between the BSW-type and Schnittman-type collisional process. Let us extend this argument to 
our more complicated 
case of equatorial charged particles. 

Equation~\eqref{C3zeroE} cannot have a maximum on account of Eq.~\eqref{ergming}, and thus the upper bound on $E_3$ in the out$-$ region must be its value for one of the osculation points. 
Picking the higher of the values in Eq.~\eqref{oscE}, we can write the bound as follows:
\rov{E_3<\left|\frac{\omega_\mrm{H}\hat\phi-\hat\omega\phi_\mrm{H}}{\hat\phi+\hat\omega A_\varphi^\mrm{H}}\ri|\sqrt{g_{\varphi\varphi}^\mrm{H}\(\msi A_1^2-m_3^2\)}+\frac{\mca N_\mrm{H}A_t^\mrm{H}\msi A_1}{\hat\phi+\hat\omega A_\varphi^\mrm{H}}\rvt}{}
We shall maximize the bound with respect to all possible
 parameters in order to derive an unconditional bound in terms of $E_1$. First, we consider $m_3\ll\msi A_1$, which also allows us to factor out $\msi A_1$,
\rov{E_3<\(\left|\frac{\omega_\mrm{H}\hat\phi-\hat\omega\phi_\mrm{H}}{\hat\phi+\hat\omega A_\varphi^\mrm{H}}\ri|\sqrt{g_{\varphi\varphi}^\mrm{H}}+\frac{\mca N_\mrm{H}A_t^\mrm{H}}{\hat\phi+\hat\omega A_\varphi^\mrm{H}}\)\msi A_1\rvt}{E3maxOUT-m3zero}
Second, we shall express $\msi A_1$ 
using $E_1$ and maximize it with respect to other parameters of particle $1$. We can use \eqref{nhparinv} with $\mca X_\mrm{H}=0$ (as \eqref{oscE} lies on \eqref{admhypxlam}) to rewrite $\chi_1$
in terms of $E_1$ and $\lambda_1$,
\rov{\chi_1=E_1\frac{\hat\phi+\hat\omega A_\varphi^\mrm{H}}{A_t^\mrm{H}}-\lambda_1\frac{\omega_\mrm{H}\hat\phi-\hat\omega\phi_\mrm{H}}{A_t^\mrm{H}}\rvt}{x1E1lam1}
Note that gauge condition \eqref{ergming} implies that the coefficient at $E_1$ is positive.
In the $\left|\lambda_1\ri|\to\infty$ limit, for 
fixed $E_1$, using Eq.~\eqref{x1E1lam1} 
in Eq.~\eqref{A1}, we find that 
the leading order of $\msi A_1$ is
\rov{\msi A_1\approx-\lambda_1\frac{\omega_\mrm{H}\hat\phi-\hat\omega\phi_\mrm{H}}{\mca N_\mrm{H}A_t^\mrm{H}}+\sigma_1\left|\lambda_1\ri|\sqrt{\(\frac{\omega_\mrm{H}\hat\phi-\hat\omega\phi_\mrm{H}}{\mca N_\mrm{H}A_t^\mrm{H}}\)^2-\frac{1}{g_{\varphi\varphi}^\mrm{H}}}\rvt}{}
One can see that $\msi A_1$
is not real due to Eq.~\eqref{ergextg}. Therefore,
for a given $E_1$, the parameter $\msi A_1$ will 
lie in the real numbers only for
 a finite 
interval of values of $\lambda_1$, and $\pder{\msi A_1}{\lambda_1}$ will blow up with opposite signs at the opposite ends of that interval. Thus, 
 there will always 
 be an extremum with respect to $\lambda_1$. (See the Appendix for details.)

For the BSW-type process, $\sigma_1=-1$, there  will be a minimum. Hence, we shall start with the following inequality (see Eq.~\eqref{A1}):
\rov{\mca N_\mrm{H}\msi A_1\f(E_1,\lambda_1,m_1\)\leqslant \chi_1\f(E_1,\lambda_1\)\rvt}{}
In order to maximize $\chi_1\f(E_1,\lambda_1\)$ of Eq.~\eqref{x1E1lam1}, 
we need to look at values of $\lambda_1$ that satisfy
\rov{\mca N_\mrm{H}\msi A_1\f(E_1,\lambda_1,m_1\)=\chi_1\f(E_1,\lambda_1\)\rvc}{A1end}
i.e., the ends of the interval mentioned above, and on their dependence on $m_1$ (cf. the Appendix). 
The resulting unconditional bound on $\msi A_1$ with $\sigma_1=-1$ is
\rov{\msi A_1\leqslant E_1\frac{\left|\hat\phi+\hat\omega A_\varphi^\mrm{H}\ri|}{\mca N_\mrm{H}\left|A_t^\mrm{H}\ri|-\sqrt{g_{\varphi\varphi}^\mrm{H}}\left|\omega_\mrm{H}\hat\phi-\hat\omega\phi_\mrm{H}\ri|}\rvt}{A1ucbBSW}
In combination with Eq.~\eqref{E3maxOUT-m3zero},
Eq.~\eqref{A1ucbBSW}
gives us the unconditional upper bound on energy $E_3$ of a particle produced in the out$-$ regime in the BSW-type process as follows:
\rov{E_3<E_1\frac{\mca N_\mrm{H}\left|A_t^\mrm{H}\ri|+\sqrt{g_{\varphi\varphi}^\mrm{H}}\left|\omega_\mrm{H}\hat\phi-\hat\omega\phi_\mrm{H}\ri|}{\mca N_\mrm{H}\left|A_t^\mrm{H}\ri|-\sqrt{g_{\varphi\varphi}^\mrm{H}}\left|\omega_\mrm{H}\hat\phi-\hat\omega\phi_\mrm{H}\ri|}\rvt}{}

For the Schnittman-type process, $\sigma_1=+1$, we can see that we need to put $m_1=0$ to maximize $\msi A_1$. Then we can find the maximum of $\msi A_1$ with respect to $\lambda_1$ (using \eqref{A1} with \eqref{x1E1lam1}; see also the Appendix)
  and derive 
    the unconditional bound on $\msi A_1$,
\rov{\msi A_1\leqslant2E_1\frac{\mca N_\mrm{H}\(\hat\phi+\hat\omega A_\varphi^\mrm{H}\)A_t^\mrm{H}}{{\mca N}_\mrm{H}^2\(A_t^\mrm{H}\)^2-g_{\varphi\varphi}^\mrm{H}\(\omega_\mrm{H}\hat\phi-\hat\omega\phi_\mrm{H}\)^2}\rvt}{A1ucbSch}
Combining with Eq.~\eqref{E3maxOUT-m3zero}, we conclude that the unconditional upper bound on
the energy $E_3$ of a particle produced in the out$-$ regime in the Schnittman-type process is 
\rov{E_3<2E_1\frac{\mca N_\mrm{H}\left|A_t^\mrm{H}\ri|}{\mca N_\mrm{H}\left|A_t^\mrm{H}\ri|-\sqrt{g_{\varphi\varphi}^\mrm{H}}\left|\omega_\mrm{H}\hat\phi-\hat\omega\phi_\mrm{H}\ri|}\rvt}{}

Let us note that for $\omega=0$, the above results reduce to the
ones of \cite{a3}, i.e., $E_3<E_1$ for the BSW-type process and $E_3<2E_1$
for the Schnittman-type process.  The bound for the Schnittman-type process is
higher than for the BSW-type process even in the general case, due to
Eq.~\eqref{ergextg}.  On the other hand, also due to Eq.~\eqref{ergextg}, we
can see that $E_3>E_1$ is generally not prevented for the BSW-type
process, unlike in the $\omega=0$ case. However, the biggest
difference is that in the general case, the gauge-dependent factors do
not cancel, and thus the bounds need to be interpreted more carefully.

We have seen above that the collisional processes analyzed here have multiple features that were absent in the previously studied special cases. Thus, now we shall discuss how the special cases follow from the
general results.

\subsection{Special cases and the degenerate case}

\lbl{odd:specdeg}

\subsubsection{Quasiradial limit}

First, let us investigate 
how to recover the results for radially moving particles 
\cite{Zasl12c}. 
 Similarly as in Sec.~\ref{sek:admreg}, 
 we can choose 
to consider either particles that move radially with respect to 
a locally nonrotating observer 
very close to the horizon, i.e., $\lambda_3=0$, 
or particles 
that would move radially in a region devoid of the influence of dragging and of magnetic field,
i.e., $L_3=0$. However, 
 both choices 
  lead to a trivial transition, unlike in Sec.~\ref{sek:admreg}.
Considering particles with a fixed value of $\lambda_3$, 
the condition $C_3>0$, see Eq.~\eqref{plusreg}, can be restated (using Eq.~\eqref{x1E1lam1} for particle $3$) as follows:
\rov{E_3>\frac{1}{\hat\phi+\hat\omega A_\varphi^\mrm{H}}\left\{\(\omega_\mrm{H}\hat\phi-\hat\omega\phi_\mrm{H}\)\lambda_3+\frac{{\mca N}_\mrm{H}A_t^\mrm{H}}{2}\[\msi A_1+\frac{1}{\msi A_1}\(m_3^2+\frac{\lambda_3^2}{g_{\varphi\varphi}^\mrm{H}}\)\]\ri\}\rvt}{}

The key feature we want to reproduce is the existence of
a threshold value $\mu$, such that $E_3>\mu$ corresponds to the {$+$} regime and $E_3<\mu$ corresponds to the {$-$} regime. 
An indeed, by setting $\lambda_3=0$, we get a threshold value $\mu$, 
given by
\rov{\mu\equiv\frac{{\mca N}_\mrm{H}A_t^\mrm{H}}{2\(\hat\phi+\hat\omega A_\varphi^\mrm{H}\)}\(\msi A_1+\frac{m_3^2}{\msi A_1}\)\rvt}{}
Moreover, $\lambda_3=0$ lies in the out region whenever it exists. Thus, we can also see that for $\lambda_3=0$, the {heavy regime} $m_3>\msi A_1$ coincides with the in regime and the
{light regime} $m_3<\msi A_1$ with the out regime. This also 
replicates the results of \cite{Zasl12c}.

\subsubsection{Geodesic limit}

\enlargethispage{-\baselineskip}

Second, we 
discuss the transition to geodesic particles, i.e., $q_3=0$. 
We shall rewrite $C_3$ of Eq. \eqref{c3} in terms of $E_3$ and $q_3$ (using Eq.~\eqref{nhpar} and dropping the contribution proportional to $\mca X_\mrm{H}$),
\rov{C_3=-\frac{1}{\omega_\mrm{H}}\[\hat\omega E_3+q_3\(\omega_\mrm{H}\hat\phi-\hat\omega\phi_\mrm{H}\)\]-\frac{{\mca N}_\mrm{H}}{2}\[\msi A_1+\frac{1}{\msi A_1}\(m_3^2+\frac{\(E_3+q_3A_t^\mrm{H}\)^2}{g_{\varphi\varphi}^\mrm{H}\omega_\mrm{H}^2}\)\]\rvt}{C3E3q3}
(More precisely, by using Eq.~\eqref{nhpar} with \eqref{fwhornc} in the derivation of \eqref{consfin}, one can make sure that the $\mca X_\mrm{H}$ term influences only higher orders of expansion.)
The resulting expression \eqref{C3E3q3}
admits a factorization, 
\rov{C_3=-\frac{{\mca N}_\mrm{H}}{2g_{\varphi\varphi}^\mrm{H}\omega_\mrm{H}^2\msi A_1}\(E_3-\msi R_+\)\(E_3-\msi R_-\)\rvc}{}
where $\msi R_\pm$ stand for
\rov{\msi R_\pm=-q_3A_t^\mrm{H}+\frac{g_{\varphi\varphi}^\mrm{H}\msi A_1}{\mca N_\mrm{H}}\[-\,\omega_\mrm{H}\hat\omega\pm\left|\omega_\mrm{H}\ri|\sqrt{{\hat\omega}^2-\frac{2q_3\mca N_\mrm{H}}{g_{\varphi\varphi}^\mrm{H}\msi A_1}\(\hat\phi+\hat\omega A_\varphi^\mrm{H}\)-\frac{{\mca N}_\mrm{H}^2}{g_{\varphi\varphi}^\mrm{H}}\(1+\frac{m_3^2}{\msi A_1^2}\)}\]\rvt}{Rpmgen}
Since $\msi R_+>\msi R_-$, the {$+$} regime corresponds to $\msi R_-<E_3<\msi R_+$ for a fixed value of $q_3$.
Let us now rewrite $\sigma_3$ of Eq.~\eqref{sig3} 
in terms of $E_3$ and $q_3$,
\rov{\sigma_3=\sgn\f[\msi A_1^2-\(m_3^2+\frac{\(E_3+q_3A_t^\mrm{H}\)^2}{g_{\varphi\varphi}^\mrm{H}\omega_\mrm{H}^2}\)\]\rvt}{}
The result again admits a factorization,
\rov{\sigma_3=-\sgn\f[\(E_3-\msi S_+\)\(E_3-\msi S_-\)\]}{}
where $\msi S_\pm$ are 
\rov{\msi S_\pm=-q_3A_t^\mrm{H}\pm\left|\omega_\mrm{H}\ri|\sqrt{g_{\varphi\varphi}^\mrm{H}\(\msi A_1^2-m_3^2\)}\rvt}{Spmgen}
As $\msi S_+>\msi S_-$, the out regime corresponds to $\msi S_-<E_3<\msi S_+$ for a fixed value of $q_3$.

Now, let us put $q_3=0$ in the equations above to find the geodesic limit.
For geodesic particles, i.e., for $q_3=0$,
it should be possible to produce particles with high values
of $E_3$ or $m_3$ only in the in$-$ regime, which prevents their
escape. 
In putting $q_3=0$, we denote the resulting values of $\msi R_\pm$
for the geodesic limit as
$\msi R_\pm^\mrm{g}$, so that 
\rov{\msi R^\mrm{g}_\pm=\frac{g_{\varphi\varphi}^\mrm{H}\msi A_1}{\mca N_\mrm{H}}\[-\omega_\mrm{H}\hat\omega\pm\left|\omega_\mrm{H}\ri|\sqrt{{\hat\omega}^2-\frac{{\mca N}_\mrm{H}^2}{g_{\varphi\varphi}^\mrm{H}}\(1+\frac{m_3^2}{\msi A_1^2}\)}\]\rvt}{}
Since $\msi S_-$ becomes negative for $q_3=0$ and $E_3>0$, 
we need to consider only $\msi S_+$ in the geodesic limit, i.e., 
$\msi S_+^\mrm{g}$, which reads
\rov{\msi S^\mrm{g}_+=\left|\omega_\mrm{H}\ri|\sqrt{g_{\varphi\varphi}^\mrm{H}\(\msi A_1^2-m_3^2\)}\rvt}{}
 If a geodesic particle $3$ 
has sufficiently high energy, such that it satisfies both $E_3>\msi R_+^\mrm{g}$ and $E_3>\msi S_+^\mrm{g}$, it will be produced in the in$-$ regime and fall into the black hole. Conversely, we can see that $\msi R_\pm^\mrm{g}$ and $\msi S_+^\mrm{g}$ all become imaginary for $m_3\gg\msi A_1$, and thus
the in$-$ regime is the only possible regime in that case. 
Hence, we 
 reproduced the results 
  of \cite{HaNeMi, Zasl12b} that the mass and energy of escaping geodesic particles is bounded.
  Note that in \cite{HaNeMi, Zasl12b}, the symbols $\lambda _\pm$ were used for $\msi R_\pm^\mrm{g}$ and $\lambda_0$ for $\msi S_+^\mrm{g}$.

\subsubsection{Degenerate case}

Third, let us
return to the degenerate case given in Eq.~\eqref{lindeg}, which was so far excluded from our discussion of energy extraction. 
We shall use the same parametrization as in the geodesic case. If we apply Eq.~\eqref{lindeg} to $\msi R_\pm$
given in Eq.~\eqref{Rpmgen}, 
they go over to degenerate $\msi R_\pm$, i.e., $\msi R_\pm^\mrm{d}$, which read 
\rov{\msi R^\mrm{d}_\pm=-q_3A_t^\mrm{H}-\frac{g_{\varphi\varphi}^\mrm{H}\msi A_1}{\mca N_\mrm{H}}\[\omega_\mrm{H}\hat\omega\mp\left|\omega_\mrm{H}\ri|\sqrt{{\hat\omega}^2-\frac{{\mca N}_\mrm{H}^2}{g_{\varphi\varphi}^\mrm{H}}\(1+\frac{m_3^2}{\msi A_1^2}\)}\]\rvt}{}
We have 
determined in Sec.~\ref{odd:admdeg} 
 that we need to impose the gauge condition $A_t^\mrm{H}=0$ in the degenerate case. However, with this condition, it holds that $\msi R_\pm^\mrm{d}=\msi R_\pm^\mrm{g}$. 
Furthermore, putting $A_t^\mrm{H}=0$ has the same effect on $\msi S_\pm$
given in Eq.~\eqref{Spmgen} as putting $q_3=0$. 
Thus, we see that upon the gauge condition $A_t^\mrm{H}=0$, the degenerate case completely coincides with the geodesic case. 
Therefore, the degenerate case corresponds to a situation when the spacetime 
behaves locally as vacuum close to $r_\mrm{H}$. However, it can be shown that the spacetime does not 
need to be globally vacuum 
 in order for Eq.~\eqref{lindeg} 
 to be satisfied (see, e.g., \cite{Article1}).

{

\section{Results for Kerr-Newman solution}

\lbl{sek:kn}

\enlargethispage{\baselineskip}

\subsection{List of relevant quantities}

Let us now apply the framework developed above to the case of extremal Kerr-Newman solution with mass $M$, angular momentum $aM$, and charge $Q$. The extremal case is defined by the condition 
\rov{
M^2=Q^2+a^2\rvc}{}
which implies 
\rov{r_\mrm{H}=M\rvt}{}
We also list the quantities relevant for our discussion as follows:
\prov{\mca N_\mrm{H}&=\frac{\sqrt{Q^2+a^2}}{Q^2+2a^2}\rvc&g_{\varphi\varphi}^\mrm{H}&=\frac{\(Q^2+2a^2\)^2}{Q^2+a^2}\rvc\\
\omega_\mrm{H}&=\frac{a}{Q^2+2a^2}\rvc&\phi_\mrm{H}&=\frac{Q\sqrt{Q^2+a^2}}{Q^2+2a^2}\rvc\\
A_t^\mrm{H}&=-\frac{Q}{\sqrt{Q^2+a^2}}\rvc&A_\varphi^\mrm{H}&=\frac{Qa}{\sqrt{Q^2+a^2}}\rvc\\
\hat\omega&=-\frac{2a\sqrt{Q^2+a^2}}{\(Q^2+2a^2\)^2}\rvc&\hat\phi&=-\frac{Q^3}{\(Q^2+2a^2\)^2}\rvt}

\subsection{Admissible region in the parameter space}

Critical particles can approach $r=M$, whenever their parameters lie inside the admissible region in the parameter space. Equations~\eqref{qparam}-\eqref{Eparam} for the border of the admissible region go over to 
\prov{q&=\frac{\sqrt{Q^2+a^2}}{Q\(Q^2+2a^2\)}\[-2a\lambda+\sqrt{\(Q^2+2a^2\)^2m^2+\(Q^2+a^2\)\lambda^2}\]\rvc\lbl{qparamkn}\\
L&=\frac{Q^2\lambda+a\sqrt{\(Q^2+2a^2\)^2m^2+\(Q^2+a^2\)\lambda^2}}{Q^2+2a^2}\rvc\lbl{Lparamkn}\\
E&=\frac{-a\lambda+\sqrt{\(Q^2+2a^2\)^2m^2+\(Q^2+a^2\)\lambda^2}}{Q^2+2a^2}\rvt\lbl{Eparamkn}}

As we discussed in Sec.~\ref{odd:admbp}, bounds on values of $q$, $L$, and $E$ in the admissible region are given by extrema of Eqs.~\eqref{qparamkn}-\eqref{Eparamkn} as functions of $\lambda$. If $\frac{\left|a\ri|}{M}<\frac{1}{2}$, then Eq.~\eqref{qparamkn} reaches an extremum, which has a value given in Eq.~\eqref{qmin}, i.e.,
\rov{q_\mrm{b}=m\frac{\sqrt{Q^2-3a^2}}{Q}\rvc}{}
and the corresponding 
values given in Eqs.~\eqref{qbL} and \eqref{qbE} of $L$ and $E$ become
\prov{L&=\frac{ma}{\sqrt{Q^2+a^2}}\frac{3Q^2+a^2}{\sqrt{Q^2-3a^2}}\rvc&E&=\frac{m}{\sqrt{Q^2+a^2}}\frac{Q^2-a^2}{\sqrt{Q^2-3a^2}}\rvt\lbl{Eqminkn}}
For $\frac{\left|a\ri|}{M}>\frac{\sqrt{5}-1}{2}$, there exists an extremum of Eq.~\eqref{Lparamkn}, which has a value given in Eq.~\eqref{Lmin}, i.e.,
\rov{L_\mrm{b}=m\sgn a\frac{\sqrt{a^4+Q^2a^2-Q^4}}{\sqrt{Q^2+a^2}}\rvc}{}
and the corresponding values given in Eqs.~\eqref{Lbq} and \eqref{LbE} of $q$ and $E$ go over to
\prov{q&=m\frac{\left|a\ri|}{Q}\frac{3Q^2+a^2}{\sqrt{a^4+Q^2a^2-Q^4}}\rvc&E&=\frac{m\left|a\ri|}{\sqrt{Q^2+a^2}}\frac{2Q^2+a^2}{\sqrt{a^4+Q^2a^2-Q^4}}\rvt\lbl{ELminkn}}

In the standard gauge vanishing at spatial infinity, the dragging potential and the electromagnetic potential  of an extremal Kerr-Newman solution satisfy the conditions
given in Eqs.~\eqref{ergextg}
and \eqref{ergming}. Therefore, Eq.~\eqref{Eparamkn} always has a minimum, see Eq.~\eqref{Emin},
\rov{E_\mrm{min}=\frac{m\left|Q\ri|}{\sqrt{Q^2+a^2}}\rvc}{vtxen}
and the corresponding values given in Eqs.~\eqref{Eminq} and \eqref{EminL} of $q$ and $L$ turn into 
\prov{q&=\frac{m}{\left|Q\ri|}\frac{Q^2-a^2}{Q}\rvc&L&=\frac{ma}{\left|Q\ri|}\frac{2Q^2+a^2}{\sqrt{Q^2+a^2}}\rvt}

The degenerate case of Eq.~\eqref{lindeg} corresponds to the extremal Kerr solution, i.e., $a=0$. Let us note that for the extremal Reissner-Nordström solution, the border of the admissible region has a symmetry with respect to change $l\to-l$.
Such a possibility was labeled as case {\bf 3} in \cite{a2}. A summary of the general results on bounds on parameters in the admissible region is given in Table \ref{tab:sumbpgen}. 

\begin{table}[h!]

\caption{Bounds on the parameters $\tilde q$, $l$, and $\varepsilon_\mrm{cr}$ of critical particles that can approach $r=M$ in an extremal Kerr-Newman spacetime. The general case can be inferred from Eqs.~\eqref{tqmin}-\eqref{epsmin}, whereas the vacuum extremal Kerr case can be inferred from Eqs.~\eqref{lmindeg} and \eqref{epsmindeg}. (The placement of nonstrict inequalities is based on results about class II critical particles; see \cite{a2}.)
} 
\setstretch{1.75}
\begin{tabular}{w{c}{2.85cm}|w{c}{2.85cm}|w{c}{2cm}|w{c}{2.85cm}|w{c}{3.3cm}|w{c}{2cm}}
\multicolumn{2}{c|}{Kerr-Newman black hole parameters}&\multicolumn{4}{c}{Restrictions}\\
$\frac{\left|a\ri|}{M}$&
 $\frac{\left|Q\ri|}{M}$&General case&Bounds on $\tilde q$&Bounds on $l$&Bounds on $\varepsilon$\\
\hline
\hline
$1$&
$0$&Vacuum&\multirow{4}{*}{No bound}&$l\sgn a>\frac{2\left|a\ri|}{\sqrt{3}}$&$\varepsilon>\frac{1}{\sqrt{3}}$\\
\cline{1-3}\cline{5-6}
$\frac{\sqrt{5}-1}{2}<\frac{\left|a\ri|}{M}<1$&
$0<\frac{\left|Q\ri|}{M}<\sqrt{\frac{\sqrt{5}-1}{2}}$&\bf 1a2b&&$l\sgn a\geqslant\frac{\sqrt{a^4+Q^2a^2-Q^4}}{\sqrt{Q^2+a^2}}$&\multirow{6}{*}{$\varepsilon>\frac{\left|Q\ri|}{\sqrt{Q^2+a^2}}$}\\
\cline{1-3}\cline{5-5}
$\frac{\sqrt{5}-1}{2}$&
$\sqrt{\frac{\sqrt{5}-1}{2}}$&{\bf 1a2c}&&$l\sgn a>0$&\\
\cline{1-3}\cline{5-5}
$\frac{1}{2}<\frac{\left|a\ri|}{M}<\frac{\sqrt{5}-1}{2}$&
$\sqrt{\frac{\sqrt{5}-1}{2}}<\frac{\left|Q\ri|}{M}<\frac{\sqrt{3}}{2}$&\bf 1a2a&&\multirow{4}{*}{No bound}\\
\cline{1-4}
$\frac{1}{2}$&
$\frac{\sqrt{3}}{2}$&{\bf 1c2a}&$\tilde q\sgn Q>0$&\\
\cline{1-4}
$0<\frac{\left|a\ri|}{M}<\frac{1}{2}$&
$\frac{\sqrt{3}}{2}<\frac{\left|Q\ri|}{M}<1$&{\bf 1b2a}&$\tilde q\sgn Q\geqslant\frac{\sqrt{Q^2-3a^2}}{\left|Q\ri|}$&\\
\cline{1-4}
$0$&
$1$&{\bf 1b2a3}&$\tilde q\sgn Q>1$&
\end{tabular}
\lbl{tab:sumbpgen}
\end{table}

\FloatBarrier

One can observe that $E\geqslant m$ for the the values of energy given in
Eqs.~\eqref{Eqminkn} and \eqref{ELminkn}, whereas $E_\mrm{min}\leqslant m$.
This implies that the expressions, given in Eqs.~\eqref{qparamkn} and \eqref{Lparamkn}, for
the values of $q$ and $L$ on the border,
defined by Eq.~\eqref{admhypxlam}, of the admissible region, are monotonic along the part of the border corresponding to $E_\mrm{cr}\leqslant m$. Therefore, the values of $q$ and $L$ in the part of the admissible region with $E_\mrm{cr}\leqslant m$ are bounded by the values of $q$ and $L$ for points on Eq.~\eqref{admhypxlam} with $E_\mrm{cr}=m$. Using the expressions for these points obtained in \cite{a2}, we present the resulting bounds in Table \ref{tab:sumbpmb}. 

\begin{table}

\caption{
Bounds on the parameters  $\tilde q$ and $l$ of critical particles with $\varepsilon\leqslant1$ that can approach $r=M$ in an extremal Kerr-Newman spacetime. (The placement of nonstrict inequalities is based on results about class II critical particles; see \cite{a2}.)}
\setstretch{1.75}
\begin{tabular}{w{c}{2.7cm}|w{c}{2.7cm}|w{c}{6.1cm}|w{c}{3.8cm}}
\multicolumn{2}{c|}{Kerr-Newman black hole parameters}&\multicolumn{2}{c}{Restrictions for $\varepsilon\leqslant1$}\\
$\frac{\left|a\ri|}{M}$&$\frac{\left|Q\ri|}{M}$&Bounds on $\tilde q$&Bounds on $l$\\
\hline
\hline
$1$&$0$&No bound&$\frac{2\left|a\ri|}{\sqrt{3}}<l\sgn a\leqslant2\left|a\ri|$\\
\hline
$\frac{1}{\sqrt{3}}<\frac{\left|a\ri|}{M}<1$&$0<\frac{\left|Q\ri|}{M}<\sqrt{\frac{2}{3}}$&$-\frac{\sqrt{Q^2+a^2}}{\left|Q\ri|^3}\(2a^2-Q^2\)\leqslant\tilde q\sgn Q<\frac{\sqrt{Q^2+a^2}}{\left|Q\ri|}$&\multirow{3}{*}{$\left|a\ri|<l\sgn a\leqslant\left|a\ri|\frac{3Q^2+2a^2}{Q^2}$}\\
\cline{1-3}
$\frac{1}{\sqrt{3}}$&$\sqrt{\frac{2}{3}}$&$0\leqslant\tilde q\sgn Q<\sqrt{\frac{3}{2}}$\\
\cline{1-3}
$0<\frac{\left|a\ri|}{M}<\frac{1}{\sqrt{3}}$&$\sqrt{\frac{2}{3}}<\frac{\left|Q\ri|}{M}<1$&$\frac{\sqrt{Q^2+a^2}}{\left|Q\ri|^3}\(Q^2-2a^2\)\leqslant\tilde q\sgn Q<\frac{\sqrt{Q^2+a^2}}{\left|Q\ri|}$\\
\end{tabular}
\lbl{tab:sumbpmb}
\end{table}

\FloatBarrier

\subsection{Structure of the parameter space with regard to energy extraction}

As we examined in Sec.~\ref{odd:exstps}, the parameter space of nearly critical particles can be divided into 
various regions corresponding to different kinematic regimes, in which particle $3$ can be produced in our collisional process. In the extremal Kerr-Newman spacetime, Eqs.~\eqref{C3zeroq}-\eqref{C3zeroE} for the border separating the $+$ and $-$ regions become
\prov{q_3&=\frac{\sqrt{Q^2+a^2}}{Q}\left\{-\frac{2a\lambda_3}{Q^2+2a^2}+\frac{1}{2}\[\msi A_1+\frac{1}{\msi A_1}\(m_3^2+\frac{\(Q^2+a^2\)\lambda_3^2}{\(Q^2+2a^2\)^2}\)\]\ri\}\rvc\lbl{C3zeroqkn}\\
L_3&=\frac{Q^2\lambda_3}{Q^2+2a^2}+\frac{a}{2}\[\msi A_1+\frac{1}{\msi A_1}\(m_3^2+\frac{\(Q^2+a^2\)\lambda_3^2}{\(Q^2+2a^2\)^2}\)\]\rvc\lbl{C3zeroLkn}\\
E_3&=-\frac{a\lambda_3}{Q^2+2a^2}+\frac{1}{2}\[\msi A_1+\frac{1}{\msi A_1}\(m_3^2+\frac{\(Q^2+a^2\)\lambda_3^2}{\(Q^2+2a^2\)^2}\)\]\rvt\lbl{C3zeroEkn}}
The two values of $\lambda_3$ that separate in and out regions 
given in Eq.~\eqref{sig3zero} are
\rov{\lambda_3=\pm\frac{Q^2+2a^2}{\sqrt{Q^2+a^2}}\sqrt{\msi A_1^2-m_3^2}\rvt}{}

Equations~\eqref{oscq}-\eqref{oscE} for the osculation points, where
the curves given by Eqs.~\eqref{qparamkn}-\eqref{Eparamkn} and
Eqs.~\eqref{C3zeroqkn}-\eqref{C3zeroEkn} touch, turn out to be
\prov{q_3&=\frac{1}{Q}\[\sqrt{Q^2+a^2}\msi A_1\mp2a\sqrt{\msi A_1^2-m_3^2}\]\rvc\\
L_3&=a\msi A_1\pm Q^2\frac{\sqrt{\msi A_1^2-m_3^2}}{\sqrt{Q^2+a^2}}\rvc\\
E_3&=\msi A_1\mp a\frac{\sqrt{\msi A_1^2-m_3^2}}{\sqrt{Q^2+a^2}}\rvt}

As we noted in Sec.~\ref{odd:exbp}, the values of $q_3$, $L_3$, and $E_3$ in the $+$ region are always bounded.
Equations~\eqref{q3min}, \eqref{L3min}, and \eqref{E3min} for bounds on these parameters go over to
\prov{q_3^\mrm{b}&=\frac{1}{2Q}\[\frac{Q^2-3a^2}{\sqrt{Q^2+a^2}}\msi A_1+\sqrt{Q^2+a^2}\frac{m_3^2}{\msi A_1}\]\rvc\\
L_3^\mrm{b}&=\frac{1}{2}\[\frac{a^4+Q^2a^2-Q^4}{a\(Q^2+a^2\)}\msi A_1+a\frac{m_3^2}{\msi A_1}\]\rvc\\
E_3^\mrm{min}&=\frac{1}{2}\[\frac{Q^2}{Q^2+a^2}\msi A_1+\frac{m_3^2}{\msi A_1}\]\rvt}

The structure of the parameter space is visualized for a black hole with $\frac{a}{M}=\frac{1}{2}$ and a process with $m_3<\msi A_1$ in the left panel of Figure \ref{f:exknps} and for a black hole with $\frac{a}{M}=\frac{\sqrt{35}}{6}$ and a process with $m_3>\msi A_1$ on the right panel. There, it is also shown how special limiting cases discussed in Sec.~\ref{odd:specdeg} correspond to different sections of the parameter space.

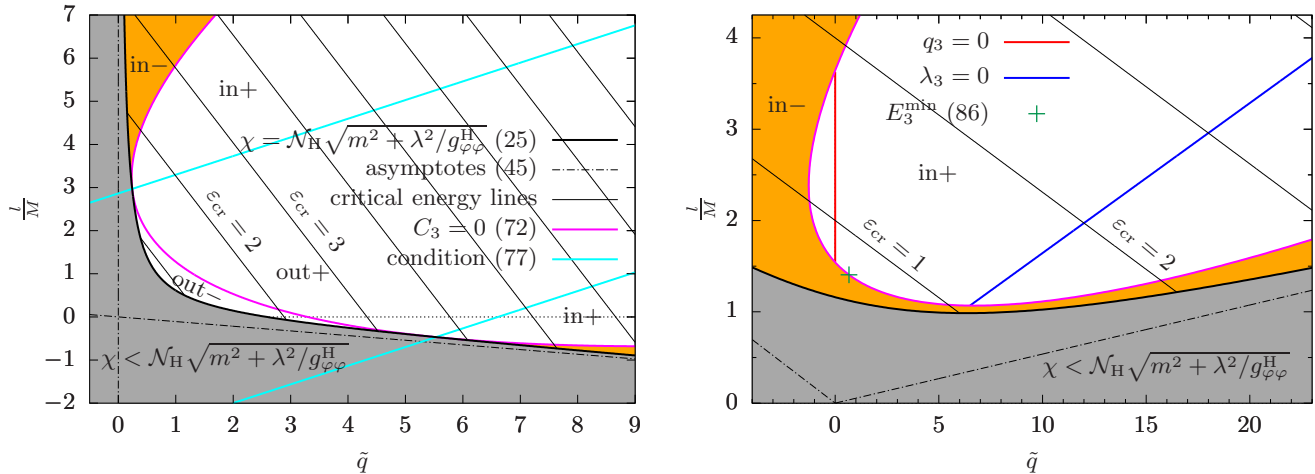
\begin{figure}
\input{exKNeqPS_ex_1k.tex}
\caption{Parameter space of critical and nearly critical particles for extremal Kerr-Newman black holes. On each panel, the part shaded in gray is outside the admissible region; i.e., critical particles with those parameters cannot approach $r=M$, and nearly critical particles produced with those parameters in the vicinity of $r=M$ cannot escape. Among the regions corresponding to different kinematic regimes of production of particle $3$ for a given $\msi A_1$, the in$-$ region, which corresponds to particle $3$ falling into the black hole, is shaded in orange. {\bf Left:} process with $\msi A_1=2.5m_3$, i.e., in the 
light regime, for a black hole with $\frac{a}{M}=\frac{1}{2}$. {\bf Right:} process with $3\msi A_1=2m_3$, i.e., in the 
heavy regime, for a black hole with $\frac{a}{M}=\frac{\sqrt{35}}{6}$. For uncharged particles in the in$+$ region (marked by the red bar), there exists a lower and an upper bound on their energy, whereas for particles with $\lambda_3=0$ in the in$+$ region (marked by the blue bar), there is only a lower bound. The ultimate lower bound on energy for the whole in$+$ region is marked by the green cross.}
\lbl{f:exknps}
\end{figure}

\section{Conclusions}

\lbl{sek:concl}

\enlargethispage{-\baselineskip}

We have studied high-energy collisions of equatorial charged test particles near the horizon of an extremal rotating electrovacuum black hole, i.e., the generalized BSW effect. Such collisions are only possible when critical particles can reach the vicinity of the horizon. Consequently, we distinguished different variants of the process based on whether there exist bounds on the charge and the angular momentum of the critical particles that can approach the horizon. (Since the values of angular momentum measured at the horizon are bounded solely in the vacuum case, we used the angular momentum measured at infinity as our reference.) Geometrically, these bounds can be seen as extrema of a hyperbola curve in the two-dimensional parameter space of the critical particles.

We proceeded to examine the possibilities of energy extraction from the black hole by particles produced in the BSW collisions, using a $2\to2$ model process. We first discussed which kinematic regimes imply escape of one of the produced particles and noted that this picture is model independent. Then, we investigated to what regions in the parameter space do these kinematic regimes correspond, and we found several situations that were not possible in the previously studied cases with fewer parameters. We also explained how these limiting cases can be obtained as different sections of the full parameter space. This is visualized in Fig.~\ref{f:exknps} for the case of extremal Kerr-Newman solution. Leaving the technical intricacies aside, one main result stands out: there are no unconditional bounds on the extracted energy as long as both the black hole and the escaping particles are charged. (And the same actually holds for the absence of unconditional upper bounds on the mass of the escaping particle.) Thus, the influence of the electromagnetic field on the energy extraction is more important in the present setup, as it prevails for arbitrarily small black hole charge. 

Although these results are very promising, we have to acknowledge that a lot of effects have not been taken into account in our considerations, most notably the electromagnetic backreaction. Additionally, as shown in \cite{a3}, despite the lack of unconditional bounds, there can be caveats which make the energy extraction unfeasible even within the test particle approximation. Although we leave the details for future work, we can nevertheless show by inspection that several of the concerns mentioned in \cite{a3} for the axial case do not apply to the equatorial case. In particular, some of the difficulties in the axial case arose from the fact that critical particles needed to be highly relativistic, $E\gg m$, in order to approach
the horizon for $Q\ll M$. 
 However, in the present case, we have shown that nonrelativistic critical particles can always approach the horizon, due to $E_\mrm{min}\leqslant m$, regardless of the value of $Q$.
 Other problems in the axial case were caused by the fact that the escaping particle $3$ needed to have a charge of larger magnitude than the initial charged critical particle $1$, i.e. $\left|q_3\ri|>\left|q_1\ri|>0$, in order to have $E_3>E_1$. In the present setup, this is not an issue whenever the black hole is rotating; in that case, the energy of critical or nearly critical particles does not need to be proportional or approximately proportional to their charge. This suggests that the presence of the frame dragging is important for the energy extraction as well, as we shall examine in follow-up work.

\section*{Acknowledgments}

F. H. thanks Jiří Bičák for introducing him to this topic and for guidance in previous research. He is also grateful to David Hilditch and Sabir Ramazanov for useful suggestions. The work of F. H. is supported by the Czech Science Foundation GAČR, Project No.~20-16531Y. F. H. would also like to acknowledge earlier support from Fundação para a Ciência e a Tecnologia (FCT), Grant No. PD/BD/113477/2015, awarded in the IDPASC framework. J. P. S. L. is grateful for the support from FCT provided through Project No.~UIDB/00099/2020. O. B. Z. thanks Kazan Federal University for a state grant for scientific activities.

\appendix

\section{Auxiliary formulas}

Let us introduce an expression $\msi P$ as follows:
\rov{\msi P=\frac{-g_{\varphi\varphi}^\mrm{H}\(\hat\phi+\hat\omega A_\varphi^\mrm{H}\)\(\omega_\mrm{H}\hat\phi-\hat\omega\phi_\mrm{H}\)E_1+\delta\sqrt{\msi W}}{{\mca N}^2_\mrm{H}\(A_t^\mrm{H}\)^2-g_{\varphi\varphi}^\mrm{H}\(\omega_\mrm{H}\hat\phi-\hat\omega\phi_\mrm{H}\)^2}\rvt}{}
We can define $\msi P_\pm$ by putting $\delta=\pm1$ and $\msi W=\msi W_\mrm{end}$, where $\msi W_\mrm{end}$ is given by
\rov{\msi W_\mrm{end}={\mca N}^2_\mrm{H}g_{\varphi\varphi}^\mrm{H}\(A_t^\mrm{H}\)^2\left\{\(\hat\phi+\hat\omega A_\varphi^\mrm{H}\)^2E_1^2-m_1^2\[{\mca N}^2_\mrm{H}\(A_t^\mrm{H}\)^2-g_{\varphi\varphi}^\mrm{H}\(\omega_\mrm{H}\hat\phi-\hat\omega\phi_\mrm{H}\)^2\]\ri\}\rvt}{}
Then, it holds that $\msi A_1\f(E_1,\lambda_1,m_1\)$ is real whenever $\lambda_1\in\[\msi P_-,\msi P_+\]$. Note that $\msi A_1\f(E_1,\lambda_1,m_1\)$ is defined by \eqref{A1} with \eqref{x1E1lam1} and also that $\lambda_1=\msi P_\pm$ satisfies Eq.~\eqref{A1end}, i.e., the square root in Eq.~\eqref{A1} being equal to zero.
It is possible to further define $\msi P_\mrm{ex}$ by putting $\delta=-\sigma_1\sgn\f[\(A_t^\mrm{H}\)\(\omega_\mrm{H}\hat\phi-\hat\omega\phi_\mrm{H}\)\]$ and $\msi W=\msi W_\mrm{ex}$, where $\msi W_\mrm{ex}$ has the following relation to $\msi W_\mrm{end}$:
\rov{\msi W_\mrm{ex}=\frac{g_{\varphi\varphi}^\mrm{H}}{{\mca N}^2_\mrm{H}}\(\frac{\omega_\mrm{H}\hat\phi-\hat\omega\phi_\mrm{H}}{A_t^\mrm{H}}\)^2\msi W_\mrm{end}\rvt}{}
One can make sure that $\msi A_1\f(E_1,\lambda_1,m_1\)$ reaches an extremum with respect to $\lambda_1$ for $\lambda_1=\msi P_\mrm{ex}$.

\end{document}

%% file: exKNeqPS_ex_1k.tex
\begingroup
  \makeatletter
  \providecommand\color[2][]{%
    \GenericError{(gnuplot) \space\space\space\@spaces}{%
      Package color not loaded in conjunction with
      terminal option `colourtext'%
    }{See the gnuplot documentation for explanation.%
    }{Either use 'blacktext' in gnuplot or load the package
      color.sty in LaTeX.}%
    \renewcommand\color[2][]{}%
  }%
  \providecommand\includegraphics[2][]{%
    \GenericError{(gnuplot) \space\space\space\@spaces}{%
      Package graphicx or graphics not loaded%
    }{See the gnuplot documentation for explanation.%
    }{The gnuplot epslatex terminal needs graphicx.sty or graphics.sty.}%
    \renewcommand\includegraphics[2][]{}%
  }%
  \providecommand\rotatebox[2]{#2}%
  \@ifundefined{ifGPcolor}{%
    \newif\ifGPcolor
    \GPcolorfalse
  }{}%
  \@ifundefined{ifGPblacktext}{%
    \newif\ifGPblacktext
    \GPblacktexttrue
  }{}%
  \let\gplgaddtomacro\g@addto@macro
  \gdef\gplbacktext{}%
  \gdef\gplfronttext{}%
  \makeatother
  \ifGPblacktext
    \def\colorrgb#1{}%
    \def\colorgray#1{}%
  \else
    \ifGPcolor
      \def\colorrgb#1{\color[rgb]{#1}}%
      \def\colorgray#1{\color[gray]{#1}}%
      \expandafter\def\csname LTw\endcsname{\color{white}}%
      \expandafter\def\csname LTb\endcsname{\color{black}}%
      \expandafter\def\csname LTa\endcsname{\color{black}}%
      \expandafter\def\csname LT0\endcsname{\color[rgb]{1,0,0}}%
      \expandafter\def\csname LT1\endcsname{\color[rgb]{0,1,0}}%
      \expandafter\def\csname LT2\endcsname{\color[rgb]{0,0,1}}%
      \expandafter\def\csname LT3\endcsname{\color[rgb]{1,0,1}}%
      \expandafter\def\csname LT4\endcsname{\color[rgb]{0,1,1}}%
      \expandafter\def\csname LT5\endcsname{\color[rgb]{1,1,0}}%
      \expandafter\def\csname LT6\endcsname{\color[rgb]{0,0,0}}%
      \expandafter\def\csname LT7\endcsname{\color[rgb]{1,0.3,0}}%
      \expandafter\def\csname LT8\endcsname{\color[rgb]{0.5,0.5,0.5}}%
    \else
      \def\colorrgb#1{\color{black}}%
      \def\colorgray#1{\color[gray]{#1}}%
      \expandafter\def\csname LTw\endcsname{\color{white}}%
      \expandafter\def\csname LTb\endcsname{\color{black}}%
      \expandafter\def\csname LTa\endcsname{\color{black}}%
      \expandafter\def\csname LT0\endcsname{\color{black}}%
      \expandafter\def\csname LT1\endcsname{\color{black}}%
      \expandafter\def\csname LT2\endcsname{\color{black}}%
      \expandafter\def\csname LT3\endcsname{\color{black}}%
      \expandafter\def\csname LT4\endcsname{\color{black}}%
      \expandafter\def\csname LT5\endcsname{\color{black}}%
      \expandafter\def\csname LT6\endcsname{\color{black}}%
      \expandafter\def\csname LT7\endcsname{\color{black}}%
      \expandafter\def\csname LT8\endcsname{\color{black}}%
    \fi
  \fi
    \setlength{\unitlength}{0.0500bp}%
    \ifx\gptboxheight\undefined%
      \newlength{\gptboxheight}%
      \newlength{\gptboxwidth}%
      \newsavebox{\gptboxtext}%
    \fi%
    \setlength{\fboxrule}{0.5pt}%
    \setlength{\fboxsep}{1pt}%
\begin{picture}(10204.00,3684.00)%
    \gplgaddtomacro\gplbacktext{%
      \csname LTb\endcsname
      \put(558,576){\makebox(0,0)[r]{\strut{}$-2$}}%
      \put(558,901){\makebox(0,0)[r]{\strut{}$-1$}}%
      \put(558,1226){\makebox(0,0)[r]{\strut{}$0$}}%
      \put(558,1552){\makebox(0,0)[r]{\strut{}$1$}}%
      \put(558,1877){\makebox(0,0)[r]{\strut{}$2$}}%
      \put(558,2202){\makebox(0,0)[r]{\strut{}$3$}}%
      \put(558,2527){\makebox(0,0)[r]{\strut{}$4$}}%
      \put(558,2853){\makebox(0,0)[r]{\strut{}$5$}}%
      \put(558,3178){\makebox(0,0)[r]{\strut{}$6$}}%
      \put(558,3503){\makebox(0,0)[r]{\strut{}$7$}}%
      \put(882,396){\makebox(0,0){\strut{}$0$}}%
      \put(1315,396){\makebox(0,0){\strut{}$1$}}%
      \put(1748,396){\makebox(0,0){\strut{}$2$}}%
      \put(2181,396){\makebox(0,0){\strut{}$3$}}%
      \put(2613,396){\makebox(0,0){\strut{}$4$}}%
      \put(3046,396){\makebox(0,0){\strut{}$5$}}%
      \put(3479,396){\makebox(0,0){\strut{}$6$}}%
      \put(3912,396){\makebox(0,0){\strut{}$7$}}%
      \put(4344,396){\makebox(0,0){\strut{}$8$}}%
      \put(4777,396){\makebox(0,0){\strut{}$9$}}%
      \csname LTb\endcsname
      \put(1575,2170){\rotatebox{-52}{\makebox(0,0)[l]{\strut{}$\varepsilon_\mrm{cr}=2$}}}%
      \put(2202,2170){\rotatebox{-52}{\makebox(0,0)[l]{\strut{}$\varepsilon_\mrm{cr}=3$}}}%
      \put(2072,1552){\makebox(0,0)[l]{\strut{}out$+$}}%
      \put(1293,1519){\rotatebox{-25}{\makebox(0,0)[l]{\strut{}out$-$}}}%
      \put(1640,2934){\makebox(0,0)[l]{\strut{}in$+$}}%
      \put(4236,1226){\makebox(0,0)[l]{\strut{}in$+$}}%
    }%
    \gplgaddtomacro\gplfronttext{%
      \csname LTb\endcsname
      \put(171,2039){\rotatebox{-270}{\makebox(0,0){\strut{}$\frac{l}{M}$}}}%
      \put(2721,126){\makebox(0,0){\strut{}$\tilde q$}}%
      \csname LTb\endcsname
      \put(4044,2561){\makebox(0,0)[r]{\strut{}$\chi=\mca N_\mrm{H}\sqrt{m^2+{\lambda^2}/{g_{\varphi\varphi}^\mrm{H}}}$ \eqref{admhypxlam}}}%
      \csname LTb\endcsname
      \put(4044,2336){\makebox(0,0)[r]{\strut{}asymptotes \eqref{admhypaskn}}}%
      \csname LTb\endcsname
      \put(4044,2111){\makebox(0,0)[r]{\strut{}critical energy lines}}%
      \csname LTb\endcsname
      \put(4044,1886){\makebox(0,0)[r]{\strut{}$C_3=0$ \eqref{C3zeroxlam}}}%
      \csname LTb\endcsname
      \put(4044,1661){\makebox(0,0)[r]{\strut{}condition \eqref{sig3zero}}}%
      \csname LTb\endcsname
      \put(558,576){\makebox(0,0)[r]{\strut{}$-2$}}%
      \put(558,901){\makebox(0,0)[r]{\strut{}$-1$}}%
      \put(558,1226){\makebox(0,0)[r]{\strut{}$0$}}%
      \put(558,1552){\makebox(0,0)[r]{\strut{}$1$}}%
      \put(558,1877){\makebox(0,0)[r]{\strut{}$2$}}%
      \put(558,2202){\makebox(0,0)[r]{\strut{}$3$}}%
      \put(558,2527){\makebox(0,0)[r]{\strut{}$4$}}%
      \put(558,2853){\makebox(0,0)[r]{\strut{}$5$}}%
      \put(558,3178){\makebox(0,0)[r]{\strut{}$6$}}%
      \put(558,3503){\makebox(0,0)[r]{\strut{}$7$}}%
      \put(882,396){\makebox(0,0){\strut{}$0$}}%
      \put(1315,396){\makebox(0,0){\strut{}$1$}}%
      \put(1748,396){\makebox(0,0){\strut{}$2$}}%
      \put(2181,396){\makebox(0,0){\strut{}$3$}}%
      \put(2613,396){\makebox(0,0){\strut{}$4$}}%
      \put(3046,396){\makebox(0,0){\strut{}$5$}}%
      \put(3479,396){\makebox(0,0){\strut{}$6$}}%
      \put(3912,396){\makebox(0,0){\strut{}$7$}}%
      \put(4344,396){\makebox(0,0){\strut{}$8$}}%
      \put(4777,396){\makebox(0,0){\strut{}$9$}}%
      \put(969,3113){\makebox(0,0)[l]{\strut{}in$-$}}%
      \put(753,917){\makebox(0,0)[l]{\strut{}$\chi<\mca N_\mrm{H}\sqrt{m^2+{\lambda^2}/{g_{\varphi\varphi}^\mrm{H}}}$}}%
    }%
    \gplgaddtomacro\gplbacktext{%
      \csname LTb\endcsname
      \put(5552,576){\makebox(0,0)[r]{\strut{}$0$}}%
      \put(5552,1265){\makebox(0,0)[r]{\strut{}$1$}}%
      \put(5552,1953){\makebox(0,0)[r]{\strut{}$2$}}%
      \put(5552,2642){\makebox(0,0)[r]{\strut{}$3$}}%
      \put(5552,3331){\makebox(0,0)[r]{\strut{}$4$}}%
      \put(6285,396){\makebox(0,0){\strut{}$0$}}%
      \put(7066,396){\makebox(0,0){\strut{}$5$}}%
      \put(7848,396){\makebox(0,0){\strut{}$10$}}%
      \put(8629,396){\makebox(0,0){\strut{}$15$}}%
      \put(9410,396){\makebox(0,0){\strut{}$20$}}%
      \csname LTb\endcsname
      \put(6519,1953){\rotatebox{-40}{\makebox(0,0)[l]{\strut{}$\varepsilon_\mrm{cr}=1$}}}%
      \put(8395,1953){\rotatebox{-40}{\makebox(0,0)[l]{\strut{}$\varepsilon_\mrm{cr}=2$}}}%
      \put(6910,2298){\makebox(0,0)[l]{\strut{}in$+$}}%
    }%
    \gplgaddtomacro\gplfronttext{%
      \csname LTb\endcsname
      \put(5273,2039){\rotatebox{-270}{\makebox(0,0){\strut{}$\frac{l}{M}$}}}%
      \put(7769,126){\makebox(0,0){\strut{}$\tilde q$}}%
      \csname LTb\endcsname
      \put(7449,3299){\makebox(0,0)[r]{\strut{}$q_3=0$}}%
      \csname LTb\endcsname
      \put(7449,3029){\makebox(0,0)[r]{\strut{}$\lambda_3=0$}}%
      \csname LTb\endcsname
      \put(7449,2759){\makebox(0,0)[r]{\strut{}$E_3^\mrm{min}$ \eqref{E3min}}}%
      \csname LTb\endcsname
      \put(5552,576){\makebox(0,0)[r]{\strut{}$0$}}%
      \put(5552,1265){\makebox(0,0)[r]{\strut{}$1$}}%
      \put(5552,1953){\makebox(0,0)[r]{\strut{}$2$}}%
      \put(5552,2642){\makebox(0,0)[r]{\strut{}$3$}}%
      \put(5552,3331){\makebox(0,0)[r]{\strut{}$4$}}%
      \put(6285,396){\makebox(0,0){\strut{}$0$}}%
      \put(7066,396){\makebox(0,0){\strut{}$5$}}%
      \put(7848,396){\makebox(0,0){\strut{}$10$}}%
      \put(8629,396){\makebox(0,0){\strut{}$15$}}%
      \put(9410,396){\makebox(0,0){\strut{}$20$}}%
      \put(5777,2814){\makebox(0,0)[l]{\strut{}in$-$}}%
      \put(7848,851){\makebox(0,0)[l]{\strut{}$\chi<\mca N_\mrm{H}\sqrt{m^2+{\lambda^2}/{g_{\varphi\varphi}^\mrm{H}}}$}}%
    }%
    \gplbacktext
    \put(0,0){\includegraphics{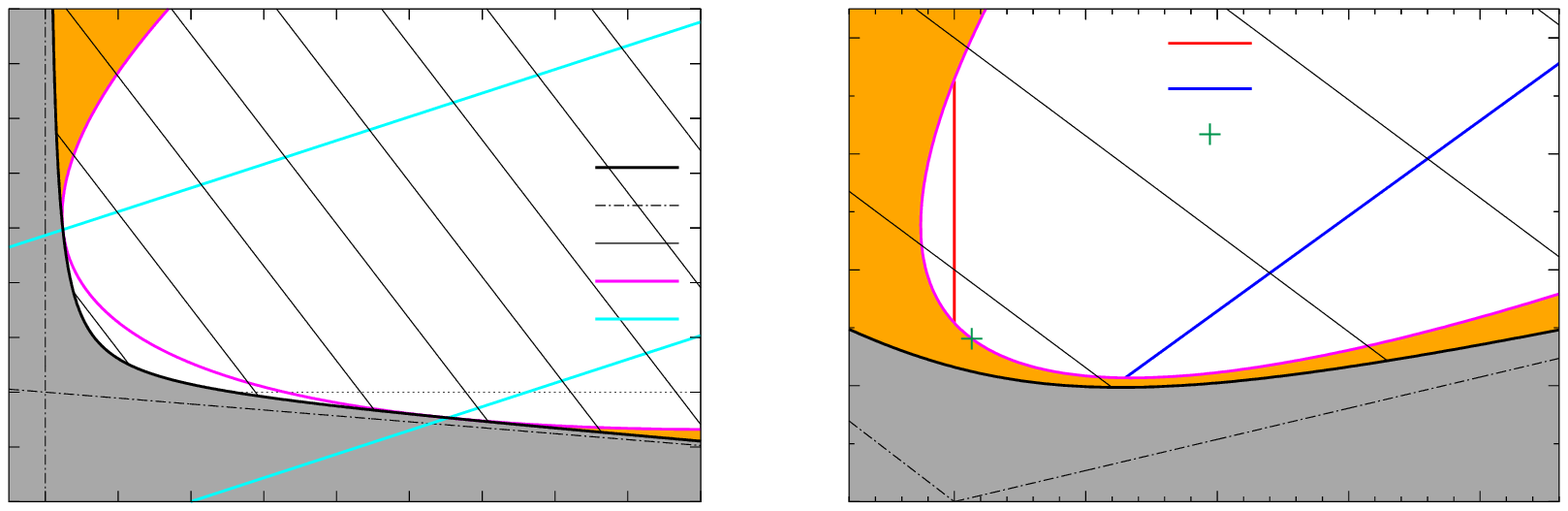}}%
    \gplfronttext
  \end{picture}%
\endgroup